\documentclass[aps,prd,twocolumn,superscriptaddress]{revtex4-1}
\usepackage{amssymb}
\usepackage[figuresright]{rotating}
\usepackage{color}
\usepackage{amsmath} 
\usepackage{amssymb} 
\usepackage{graphics}
\usepackage{xspace}
\usepackage[gen]{eurosym}
\usepackage{latexsym}
\usepackage[mathscr]{eucal} 
\usepackage{latexsym}
\usepackage[left]{lineno}

\newcommand{\nue}{\ensuremath{\nu_{e}}\xspace}
\newcommand{\nubare}{\ensuremath{\overline{\nu}_{e}}\xspace}

\newcommand{\numu}{\ensuremath{\nu_{\mu}}\xspace}

\newcommand{\nubarmu}{\ensuremath{\overline{\nu}_{\mu}}\xspace}

\newcommand{\numunue}{\ensuremath{\nu_\mu \rightarrow \nu_e}\xspace}

\newcommand{\boss}[2]{\ensuremath{\rlap{\kern-2.5pt\ensuremath{\overset{\scriptscriptstyle(-)}{\phantom{#1}}}}{\ensuremath{{#1}_{#2}}}}}

\newcommand{\bOLOGNAI}     {INFN-CNAF, 40127 Bologna, Italy}
\newcommand{\pADOVAI}          {INFN, Sezione di Padova, 35131 Padova, Italy}

\begin{document}


\title{Determination of the neutrino mass hierarchy with a new statistical method}
\vskip 10pt
\author{L.~Stanco}\affiliation{\pADOVAI}
\author{S.~Dusini}\affiliation{\pADOVAI}
\author{M.~Tenti}\affiliation{\bOLOGNAI}

\noaffiliation
\vskip10pt

\date{\today}

\vskip 20pt
\begin{abstract}
Nowadays neutrino physics is undergoing a change of perspective: the discovery period is almost over and
the phase of precise measurements is starting. Despite the limited statistics collected for some variables, the three--flavour 
oscillation neutrino framework is strengthening well. 
In this framework a new method has been developed to determine the neutrino mass ordering, one of the still
unknown and most relevant parameters. The method is applied to the 2015 results of the NOvA experiment for \numunue appearance,
including its systematic errors.
A substantial gain in significance is obtained compared to the traditional $\Delta\chi^2$  approach.
Perspectives are provided for future results obtainable by NOvA with larger exposures. 
Assuming the number of the 2015 \nue observed events scales with the exposure, an increase in only a factor three would exclude
the inverted hierarchy at more than 95\% C.L. over the full range of the CP violating phase. 
The preliminary 2016 NOvA measurement
on \numunue appearance has  also been analyzed.
\end{abstract}

\maketitle


\section{Introduction}\label{sec:intro}

The unfolding of neutrino physics is a long and pivotal history spanning the past 80 years. Over that period of time the interplay of 
theoretical hypotheses and experimental facts was one of the most fruitful to make progress in particle physics.
The achievements of the past two decades brought out a coherent picture within the Standard Model or some minor extensions of it,
namely the mixing of three neutrino flavour--states, $\nu_e$, $\nu_{\mu}$ and $\nu_{\tau}$, with three  $\nu_1$, $\nu_2$ and $\nu_3$ mass eigenstates. 
After  determining the absolute masses of neutrinos, their Majorana/Dirac nature, the existence and the magnitude of the leptonic CP violation,
the (standard) three--neutrino model will be completely settled. However, the first two questions 
will probably take some time to be answered, while the third one is a matter of debates and experimental proposals. 

Actually, in the three--neutrino framework  an unknown parameter is closely tied to the masses and
the CP violating phase, $\delta_{CP}$: the neutrino mass ordering of the neutrino mass eigenstates.
Namely, it is still largely unconstrained the sign of $\Delta m_{31}^2=m_3^2 - m_1^2$, the  difference 
of the squared masses of $\nu_3$ and $\nu_1$. Its knowledge is of utmost importance to provide inputs for future studies and
experimental proposals, to finally clarify whether we need  new projects at all, and to constrain analyses in other fields
such as cosmology and astrophysics.

The mass ordering (MO) is usually identified as normal hierarchy (NH) when $\Delta m_{31}^2 > 0$ or inverted 
hierarchy (IH) in the opposite case. 
All the methods developed so far for establishing whether MO is normal or inverted are based on $\chi^2$ evaluation.
Given the current uncertainties of the oscillation parameters~\cite{lisi2016}
from few percents to more than 10\%,
the computation of the difference of the $\chi^2$  best fits for NH and IH is performed~\cite{mh-all}. 
These analyses use the test statistic

{\footnotesize
\begin{equation}
\Delta\chi^2_{min}= \chi^2_{min}({\rm IH})-\chi^2_{min}({\rm NH}),
\end{equation}}
\noindent where the two minima are evaluated spanning the uncertainties of the three-neutrino oscillation parameters,
namely $\Delta m_{21}^2$, $\pm\Delta m_{31}^2$, $\theta_{12}$, $\theta_{23}$, $\theta_{13}$ and $\delta_{CP}$.
$\theta_{ij}$ ($i,j=1,2,3$) are the mixing angles in the standard parameterization
and $\Delta m_{21}^2=m_2^2 - m_1^2$.
The statistical significance in terms of standard deviations is  computed as $\sqrt{\Delta\chi^2}$. The limits of such procedures
are well known~\cite{ciufoli}. In particular, the significance corresponds only to the median expectation and does not consider the
intrinsic statistical fluctuations. Thus, errors of type I and II~\cite{pdg} should be taken into account when comparing the probability density functions
of each $\chi^2_{min}$
As a consequence the corrected significance is lower and more $\sigma$'s are needed to reach a robust observation. 
Despite these caveats no alternative test statistic has been  outlined so far.

Broader discussions on the $\Delta\chi^2$ test statistic and the way to approach analyses on the mass hierarchy can be found 
in section 3 of~\cite{lucas} and references therein. The MO evaluation should be performed with a change of perspective: the achievement should
focus on the rejection of the wrong hierarchy rather than the observation of the true one. Therefore, it is mandatory to introduce
new test statistics that allow this approach to distinguish between NH and IH. 
Moreover, it is important to work out a comprehensive handling of all 
future measurements on MO.
As an alternative, the use of only one experiment is mainly due to
the lack of confidence in the 3-neutrino framework and/or in the cross-correlation of the systematic errors 
among different experiments. The first concern should be targeted with specific experiments and
it should not affect the extraction of the oscillation parameters. The second concern about the systematic errors should
not avoid using one experiment as pivot and then adding information from the other ones.

This paper aims to introduce a new method that can be extensively applied to single or multiple measurements of the neutrino mass ordering.
For the time being it has been applied to the results from the NOvA experiment on \numunue appearance, in 2015~\cite{nova-nue} and 
2016~\cite{nova-prel}.
In the following sections the NOvA environment is recalled, its simulation and the application of the $\Delta\chi^2_{min}$ method are reported, 
and then the new technique is introduced.

\section{The NOvA environment}\label{sec:nova}

The predicted number of NOvA \nue oscillated events for an exposure of $2.74\times 10^{20}$
protons-on-target (p.o.t.) is about 5 and 3 in the NH and IH hypotheses, respectively, whereas a little less than 1 event is expected from the 
background (2015 NOvA conditions~\cite{nova-nue}). The number of oscillated events 
is highly dependent on $\delta_{CP}$, and to a lesser extent on 
$\theta_{23}$ and $\theta_{13}$. Dependences on $\Delta m_{21}^2$, $\Delta m_{31}^2$ and $\theta_{12}$ are minor and therefore are 
neglected in this study. 
This behaviour (of the number of expected \nue events)
has been checked and reproduced in detail by the authors using the GLoBES package~\cite{globes}, 
although it is commonly  known~\cite{dipende}. 

\begin{figure}[htb]
\includegraphics[width=8cm,height=5.5cm]{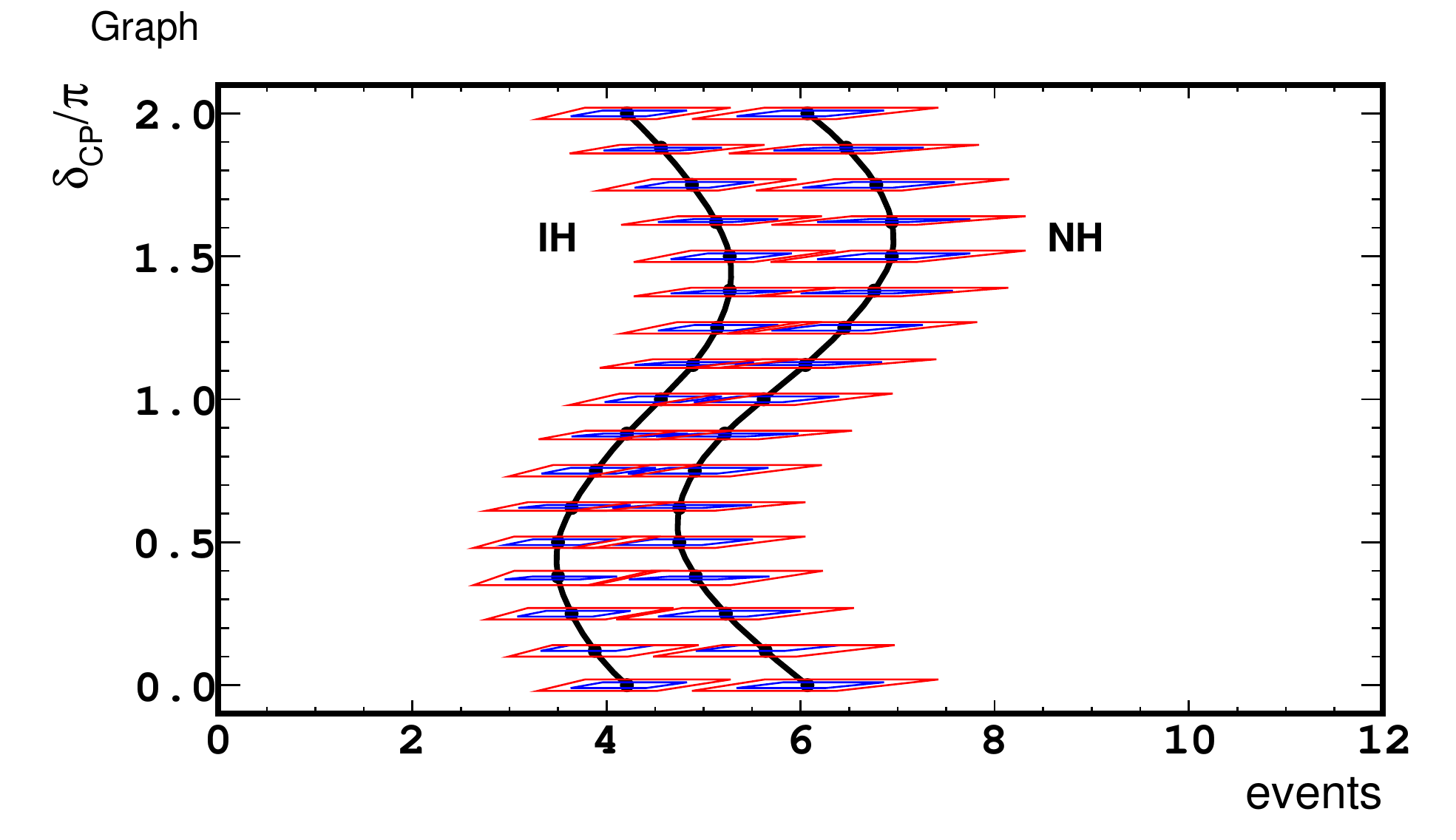}%
\caption{\label{fig1}(color online) The number of predicted oscillated \numunue events plus the expected background is shown in the horizontal axis
versus $\delta_{CP}$ in the vertical axis (the two continuous lines for the IH and NH hypotheses, respectively). Normalization is given by the 2015 NOvA analysis (LID case)~\cite{nova-scaling}, while the
neutrino oscillation parameters are taken by the best fit values of the global fit (GF) (column 3 of table 1 in~\cite{lisi2016}).
The computation has been performed with the GLoBES package. The two concentric
areas for 16 representative values of $\delta_{CP}$ spanning its range correspond to the 1 $\sigma$ and 2 $\sigma$ contours due to the (correlated)
$\theta_{23}$, $\theta_{13}$ uncertainties (see text for more explanations). The 1 $\sigma$ and 2 $\sigma$ uncertainties 
on $\theta_{23}$, $\theta_{13}$ are given 
by the GF.}
\end{figure}

In Fig.~\ref{fig1} the number of predicted oscillated \numunue events plus the expected background is shown in the  horizontal axis as function of 
$\delta_{CP}$ (vertical axis), normalized to the 2015 NOvA expectation~\cite{nova-scaling}
and taking the best fit values by the global fit (GF) in~\cite{lisi2016}.
The signal part of the predicted number of events suffers from the (correlated) uncertainties on $\theta_{23}$ and $\theta_{13}$. For each $\delta_{CP}$ value
 the predicted number of events is thus spread out due to the possible variations of $\theta_{23}$, $\theta_{13}$.
 If the estimations of the $\theta_{23}$, $\theta_{13}$ uncertainties at 1~$\sigma$ and 2~$\sigma$ levels are taken from the GF, the corresponding spreads on the number of the predicted events 
are shown as parallelograms for 16 representative values of $\delta_{CP}$ spanning its entire range.
In each parallelogram the leftmost and rightmost vertices correspond to the coherent contributions of the uncertainties (positive correlation),
[$-\delta\theta_{23}$, $-\delta\theta_{13}$] and [$+\delta\theta_{23}$, $+\delta\theta_{13}$], while the other two vertices correspond
to the counter-contributions (negative correlation) [$-\delta\theta_{23}$, $+\delta\theta_{13}$] and [$+\delta\theta_{23}$, $-\delta\theta_{13}$].
These choices are dictated by the almost linear correlations between $\theta_{23}$, $\theta_{13}$ and the \numunue 
appearance probability, in the NOvA conditions and around the best fit solutions of $\theta_{23}$, $\theta_{13}$.
The heights of the parallelograms are in arbitrary units to ensure a clear vision.

Looking at the patterns in Fig.~\ref{fig1} a conclusion is straightforward: no discrimination between IH and NH 
can be achieved if the $\chi^2$ minimization is performed in the full range of $\delta_{CP}$.
In such a kind of fit several similar solutions with $\chi^2\lesssim 1$ are possible for different values of $\delta_{CP}$. For example, 
$\chi^2_{min}({\rm NH})$ for $\delta_{CP}=0.5\,\pi$  is close to $\chi^2_{min}({\rm IH})$ for $\delta_{CP}=1.5\,\pi$.
In other words, it is always possible to find at least a couple of $\chi^2_{min}({\rm NH})$, $\chi^2_{min}({\rm IH})$ so that 
$\Delta\chi^2_{min}$ is very close to zero, i.e. IH and NH are indistinguishable.
A better discrimination between NH and IH could be obtained if minimization is performed assuming a single value of $\delta_{CP}$. 
However, the result on MO would be then closely tied to $\delta_{CP}$. Moreover, even computing $\Delta\chi^2_{min}(\delta_{CP})$
 only a
mild indication for NH is obtained, as it is shown below.

\begin{figure}
\includegraphics[width=8cm]{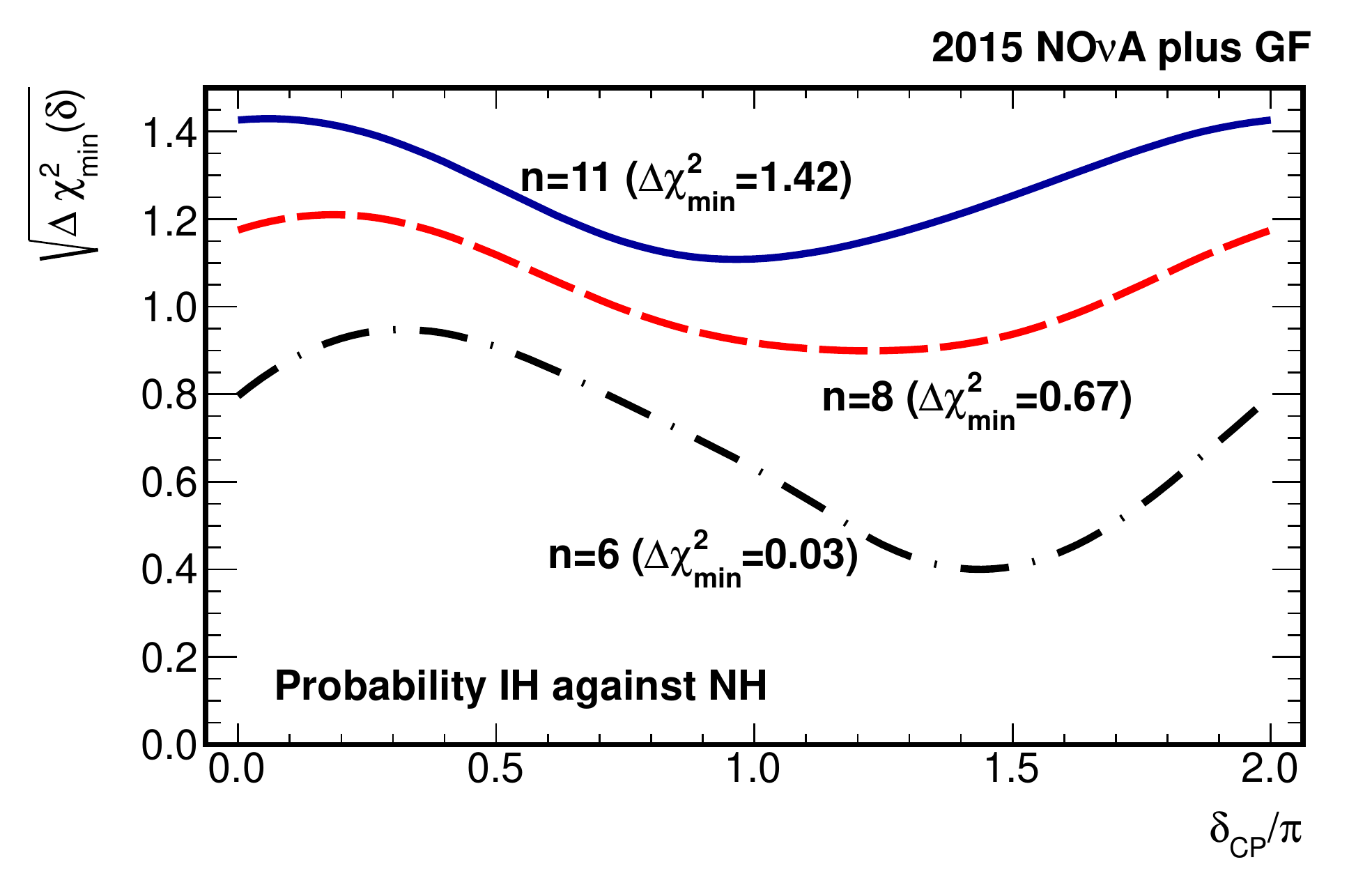}%
\caption{\label{fig2}(color online) 
The significance of IH against NH is shown as function of $\delta_{CP}$, using
the standard analysis with the $\Delta\chi^2_{min}$ test statistic.
The three curves correspond to the significances
for different numbers of events, 6, 8 or 11, observed in the 2015 NOvA analysis in the \numunue channel, for a total exposure of
$2.74\times 10^{20}$ p.o.t. The analysis has been simulated with GLoBES and normalized to the NOvA 
expectations~\cite{nova-scaling}.
The result is obtained minimizing over $\theta_{23}$, $\theta_{13}$ and using the best fit values of the global fit 
(GF)~\cite{lisi2016} for the other parameters. The quoted $\Delta\chi^2_{min}$ correspond to 
$\chi^2_{min}({\rm IH})-\chi^2_{min}({\rm NH})$ after integrating in $\delta_{CP}$.}
\end{figure}
2015 NOvA appearance result is two--fold since two different analyses were done. The primary selection technique (LID)
found 6 events, while the secondary one (LEM) found 11 events. Throughout the paper 8 events have also been considered,
as a kind of test bench. This choice is dictated by the rather low probability to observe 11 events 
compared to the NH expectation (5\% at $\delta_{CP}=0$), while 8 events have a mild, more acceptable probability (26\%). 6 events stand on the median expectation of NH at $\delta_{CP}=0$.
Using the GLoBES simulation  $\chi^2$ minimizations were made over  $\theta_{23}$, $\theta_{13}$, 
as function of $\delta_{CP}$ and for 6, 8 and 11 observed events,  to extract
$\Delta\chi^2_{min}(\delta_{CP})=\chi^2_{min}({\rm IH};\delta_{CP})-\chi^2_{min}({\rm NH};\delta_{CP})$.
In Fig.~\ref{fig2} the equivalent number of standard deviations as function of $\delta_{CP}$ is shown. 
The significances fairly reproduce what can be extracted 
by 2015 NOvA results even though the procedure is rather different (no systematic errors, different uncertainties on
$\theta_{23}$ and $\theta_{13}$ etc.)~\cite{nova-com}. 
After integrating in $\delta_{CP}$  the $\Delta\chi^2_{min}=\chi^2_{min}({\rm IH})-\chi^2_{min}({\rm NH})$ was also computed. 
For 6, 8 and 11 events small significances are obtained: 0.17, 0.82 and 1.20, respectively.
There is no doubt that an evaluation in terms of a best fit for $\Delta\chi^2$ over the full range
of $\delta_{CP}$ gives marginal results. This leads to the conclusion drawn in~\cite{lisi2016}: the sensitivity
to the mass hierarchy is currently null.

Considering what has been highlighted so far a more sophisticated test statistic should be introduced.

\section{The new test statistic}\label{sec:test}

\begin{figure}[htb]
\includegraphics[width=0.95\linewidth]{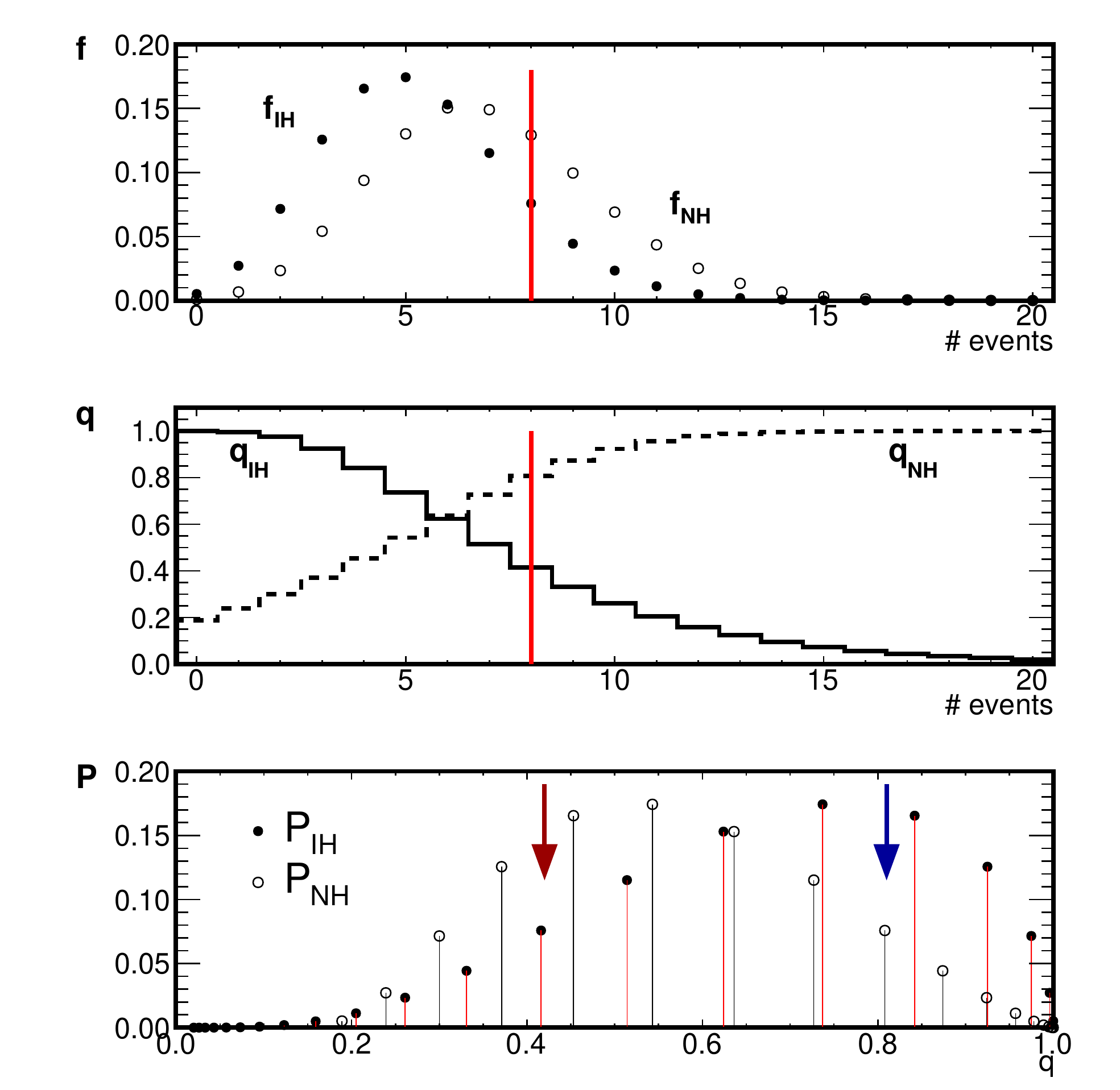}%
\caption{\label{fig3}(color online) Top: the predicted Poisson distributions of the 2015 NOvA analysis (signal plus background) are shown 
for IH (full points) and NH (open points), 
for $\delta_{CP}=3/2\, \pi$ and an exposure of $2.74\times 10^{20}$ p.o.t..
The vertical line corresponds to $n=8$.
Middle: the corresponding values assumed by $q_{\rm IH}$ (plain line) and $q_{\rm NH}$ (dashed line). 
Bottom: the probability mass functions of $q_{\rm IH}$ (full points) and $q_{\rm NH}$ (open points).
The arrows indicate the thresholds used to compute $q_{\rm IH}$ and $q_{\rm NH}$ for $n=8$. }
\end{figure}

A new test statistic $q$ is defined, following a Bayesian approach developed in a frequentist way.
For each hypothesis IH or NH one considers the Poisson distributions
$f_{\rm MO}(n_i;\mu_{\rm MO} |\delta_{CP})$, where $n_i$ is the random variable and 
$\mu_{\rm MO} (\delta_{CP} )$ is the predicted mean (signal plus background) as function of $\delta_{CP}$, 
 MO standing for IH or NH. Dependences on the oscillation parameters, in particular $\theta_{23}$, $\theta_{13}$,
 are not explicitly shown, even though they are included in the analysis.
For a specific $n$ the left and right cumulative functions of $f_{\rm IH}$ and $f_{\rm NH}$ are computed and their ratios are evaluated. The ratios are similar
to the CL$_s$ test statistic used for the Higgs discovery~\cite{cls}. Since for the \nue appearance at NOvA the expectation is asymmetric
towards IH and NH (less events are expected for IH than for NH for the \nue appearance in the \numu beam, opposite case 
holding for the \nubarmu  beam and the \nubare appearance),
the ratios $q_{\rm MO}$ are defined either for the IH or the NH case:

\vspace{-0.1cm}
 {\footnotesize
\begin{align}\label{eq2}
q_{\rm IH}(n; \delta_{CP}) & =\frac{\sum_{n_{i,{\rm IH}}\ge n} f_{\rm IH}(n_{i,{\rm IH}};\mu_{\rm IH} |\delta_{CP} )}{\sum_{n_{j,{\rm NH}}\ge n}
f_{\rm NH}(n_j({\rm NH});\mu_{\rm NH} |\delta_{CP} )},\\
\label{eq3}
q_{\rm NH}(n; \delta_{CP})&=\frac{\sum_{n_{i,{\rm NH}}\le n} f_{\rm NH}(n_{i,{\rm NH}};\mu_{\rm NH} |\delta_{CP})}{\sum_{n_{j,{\rm IH}}\le n} 
f_{\rm IH}(n_{j,{\rm IH}};\mu_{\rm IH} |\delta_{CP})}. 
\end{align}
}
$q_{\rm IH}$ and $q_{\rm NH}$ are functions of the random variable $n$~\cite{ref-note}
and therefore they are themselves two discretized random 
variables defined in the [0, 1] interval. As $n$ goes to zero  $q_{\rm IH}$
goes to one, while when $n$ increases $q_{\rm IH}$ asymptotically tends to zero. $q_{\rm NH}$ behaves the other way around 
towards $n$. For illustration purpose the behaviours of $f_{\rm MO}$ and $q_{\rm MO}$ are shown in Fig.~\ref{fig3} for a typical case ($n=8$).

The probability mass functions of $q_{\rm MO}$, $P_{\rm MO}(q_{\rm MO})$, 
were computed via toy 
Monte Carlo simulations based on $f_{\rm IH}$ (test of IH against NH) 
or $f_{\rm NH}$ (test of NH against IH).
Selecting the observed data $n_D$, the number of observed events 
either in real data or in Monte Carlo simulation, $P_{\rm MO}(q_{\rm MO})$ probabilities are used to 
evaluate the corresponding $p$--values, $p_{\rm MO}$~\cite{check}:

 {\footnotesize
\begin{align}
p_{\rm IH}(n_D; \delta_{CP})& =\sum_{q'_{\rm IH}\le q_{\rm IH}(n_D)} P_{\rm IH}(q'_{\rm IH} ;\delta_{CP}),\\
p_{\rm NH}(n_D; \delta_{CP})&= \sum_{q'_{\rm NH}\le q_{\rm NH}(n_D)} P_{\rm NH}(q'_{\rm NH} ;\delta_{CP}).
\end{align}}

Finally, the significance is computed from the $p_{\rm MO}$--values with the one--sided option.
It corresponds to 0 sigma ($Z=0$) when $p_{\rm MO}=50\%$ that equalizes the IH and NH probabilities.
Within that choice $Z$ is defined as $Z=\Phi^{-1}(1-p_{\rm MO})$, where $\Phi^{-1}$ is the quantile 
(inverse of the cumulative distribution) of the standard Gaussian and 
Z is the number of standard deviations. 
In the appendix the technical aspects of the new method are illustrated for a simplified case taking into account only the statistical errors.
A detailed comparison with $\Delta\chi^2_{min}$ results can be found too.

The dependences on $\theta_{23}$ and $\theta_{13}$ enter in the prediction of the mean $\mu_{\rm MO}$.
Their uncertainties, as well as the systematic errors evaluated for the experimental data, let fluctuate
the prediction of the median number of events.
These errors have been
taken into account using two approaches: A) convolution of the Poisson distributions with assumed Gaussian distributions~\cite{cou-high} for
the uncertainties on $\theta_{23}$, $\theta_{13}$  (central values and standard deviations being given by the GF)
and the systematic errors on signal and background  (as provided by NOvA); 
 B) evaluation of the error bands overlaying the significance, choosing a $\pm\, \sigma$ variation of
the mixing angles and the systematic errors.
 Although results will be provided for both errors' treatments  our primary choice is A for the uncertainties on $\theta_{23}$, $\theta_{13}$,
 and B for the systematic errors. In such a case  the probability distributions of $\theta_{23}$, $\theta_{13}$ are treated as a posterior
 information and used as {\em prior} for the next calculation.
Then the initial Poisson distribution $f_{\rm MO}$ becomes:

 {\footnotesize
\begin{align*}
f_{\rm MO}(n_i; & \mu_{\rm MO}  |\delta_{CP})= \int Poi_{\rm MO}(n_i;\mu_{\rm MO}(\theta_{23}',\theta_{13}')
|\delta_{CP}) \\
&\cdot G(\theta_{23}',\theta_{13}';\hat{\theta}_{23},\hat{\theta}_{13},\sigma_{\hat{\theta}_{23}}, 
\sigma_{\hat{\theta}_{13}})\,
d\theta_{23}' \, d\theta_{13}',
\end{align*}}

\noindent where $Poi$ stands for the  Poisson function and $G(\theta_{23},\theta_{13})$ is the double
Gaussian distribution centered to the best fit values $\hat{\theta}_{23}$, $\hat{\theta}_{13}$.

\section{Results}\label{sec:results}
The $q_{\rm MO}$ estimator were  applied to the \numunue appearance 2015 NOvA result first. 
\begin{figure}[h]
\includegraphics[width=7cm]{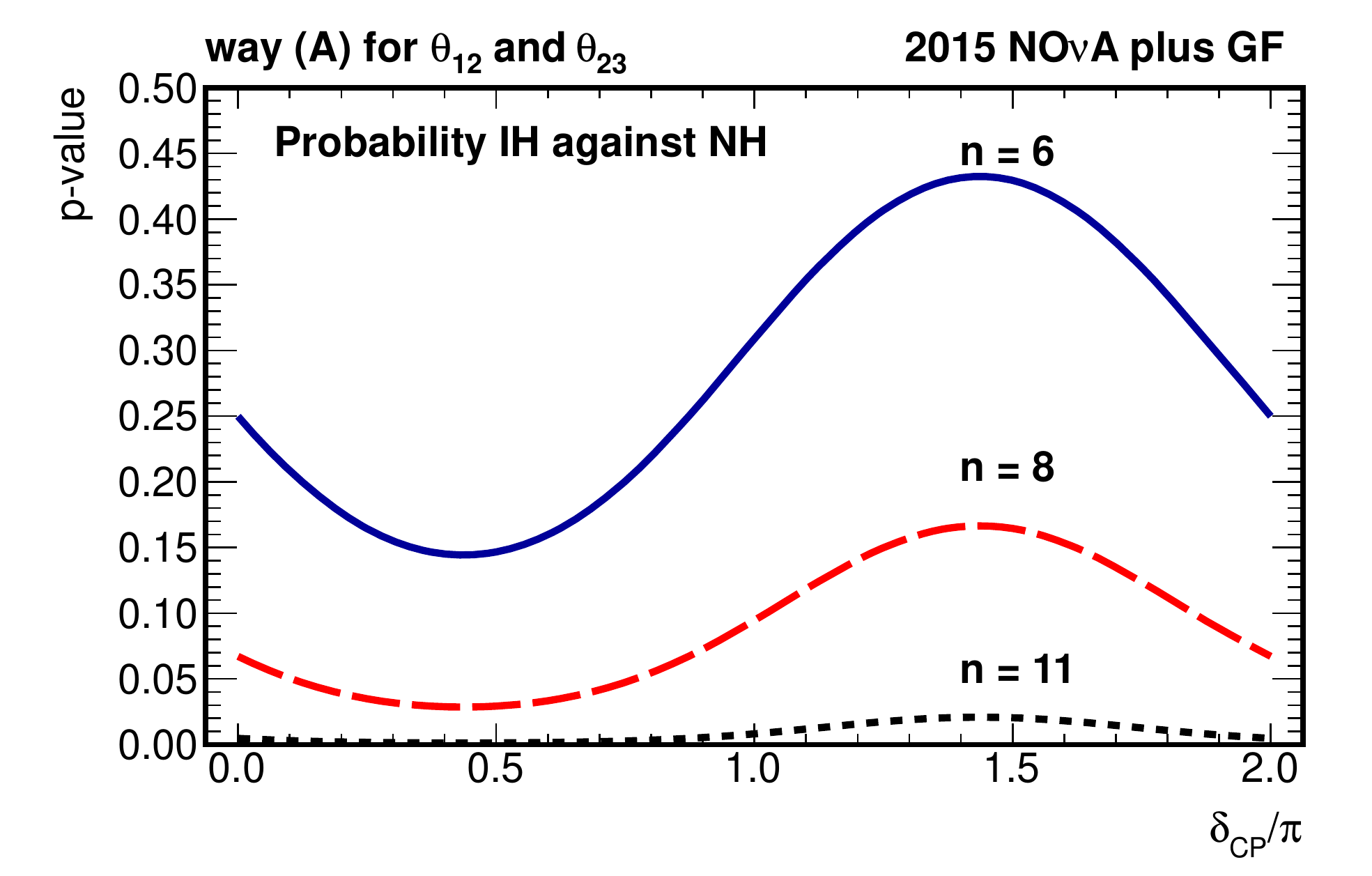}
\includegraphics[width=7cm]{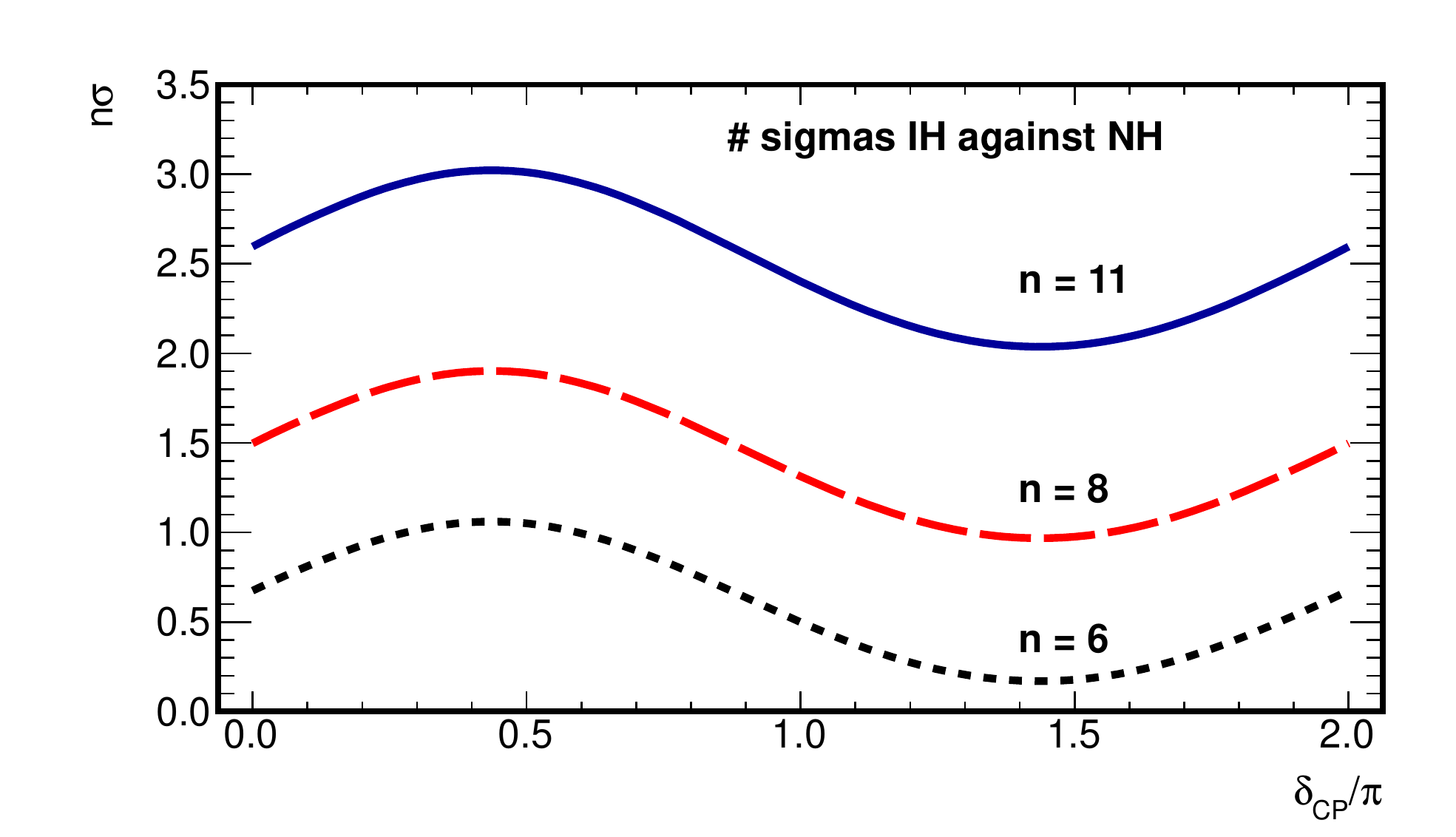}%
\caption{\label{fig4}(color online) The top plot shows the probability of IH against NH hypotheses computed by the new method 
illustrated in the text.
In the bottom picture the equivalent significance in terms of $\sigma$ (one-sided option) is drawn. The different curves correspond to
different number of events, 6, 8 and 11 (signal plus background), as measured by the 2015 NOvA condition in the \numunue channel, for a total exposure of
$2.74\times 10^{20}$ p.o.t..
}
\end{figure}
Selecting IH as truth the $p$-value probabilities and their significances are shown for $q_{\rm IH}$ in Fig.~\ref{fig4}, top and bottom, 
respectively. Results were obtained as function of $\delta_{CP}$ and for three cases: 6, 8 and 11 observed events. 
The $p$-values correspond to the 
probability to exclude IH against NH, with the oscillation parameter set given by GF. For $n_D=8$ the significances average around 
1.5~$\sigma$, with a slightly higher significance 
for $\delta_{CP}$ in $[0, \pi]$. The systematic errors have not been included, while the uncertainties on  $\theta_{23}$, $\theta_{13}$
have been treated as nuisances (approach A). 
Overall, when $n_D=8$ the new method provides an increase in 0.5 $\sigma$ compared to the $\Delta\chi^2_{min}$ method.
The increase is not constant depending on $n_D$ and $\delta_{CP}$: 
the improvement of the new method in terms of 
standard deviations  strongly increases in ``favorable'' regions of $\delta_{CP}$ and with $n_D$. For example when $n_D=11$
the increase is about 1 $\sigma$ when averaging over $\delta_{CP}$, and 1.5 $\sigma$ for $\delta_{CP}<\pi$.

The situation improves considerably if for NOvA a factor three in exposure ($8.22\times 10^{20}$ p.o.t.) is taken into account,
assuming the same 2015 efficiency of the signal and the same level of background rejection.
Compared to the $\Delta\chi^2_{min}$ method a much larger increase in significance is obtained.
\begin{figure}[h]
\includegraphics[width=7cm]{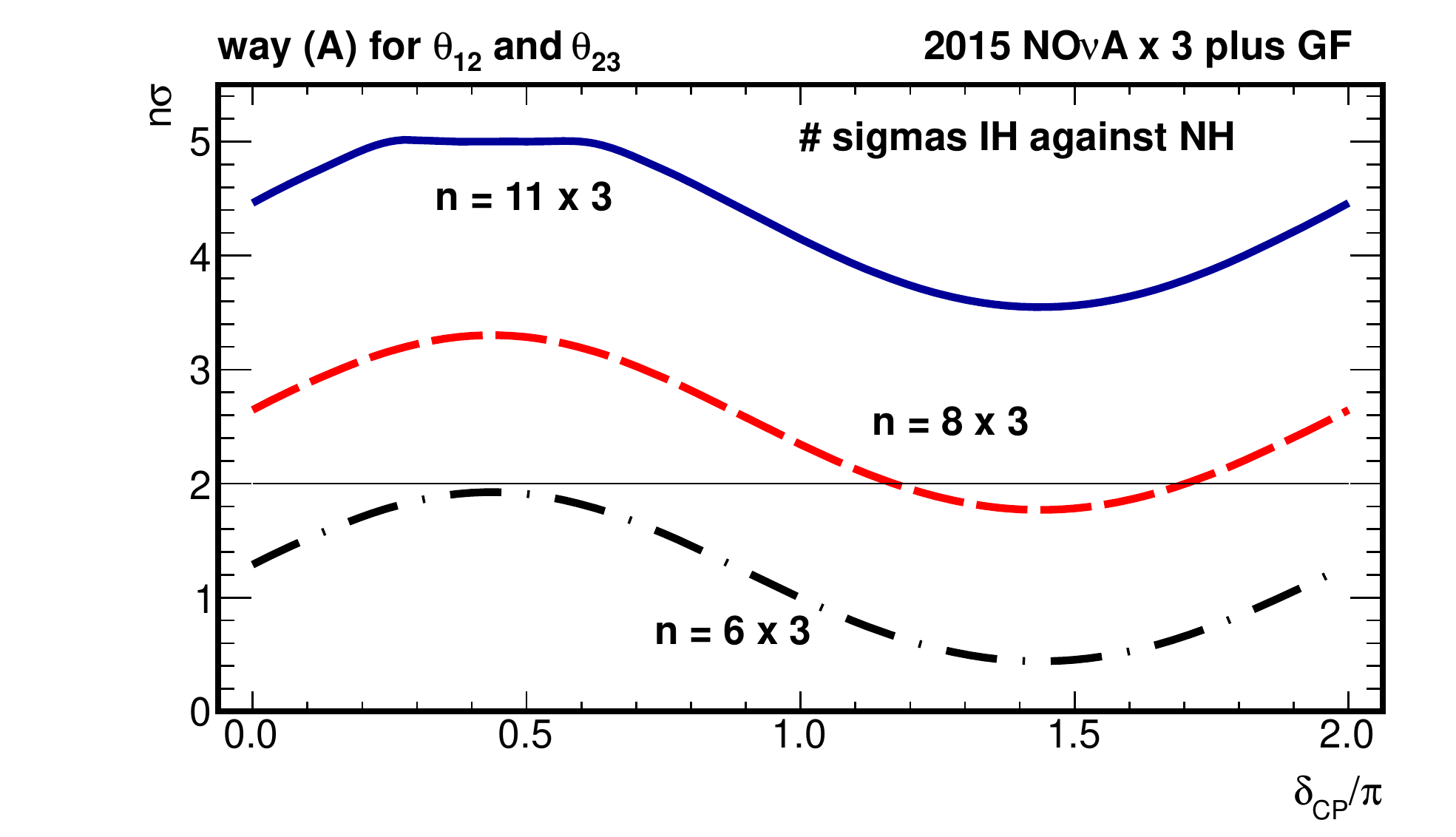}
\includegraphics[width=7cm]{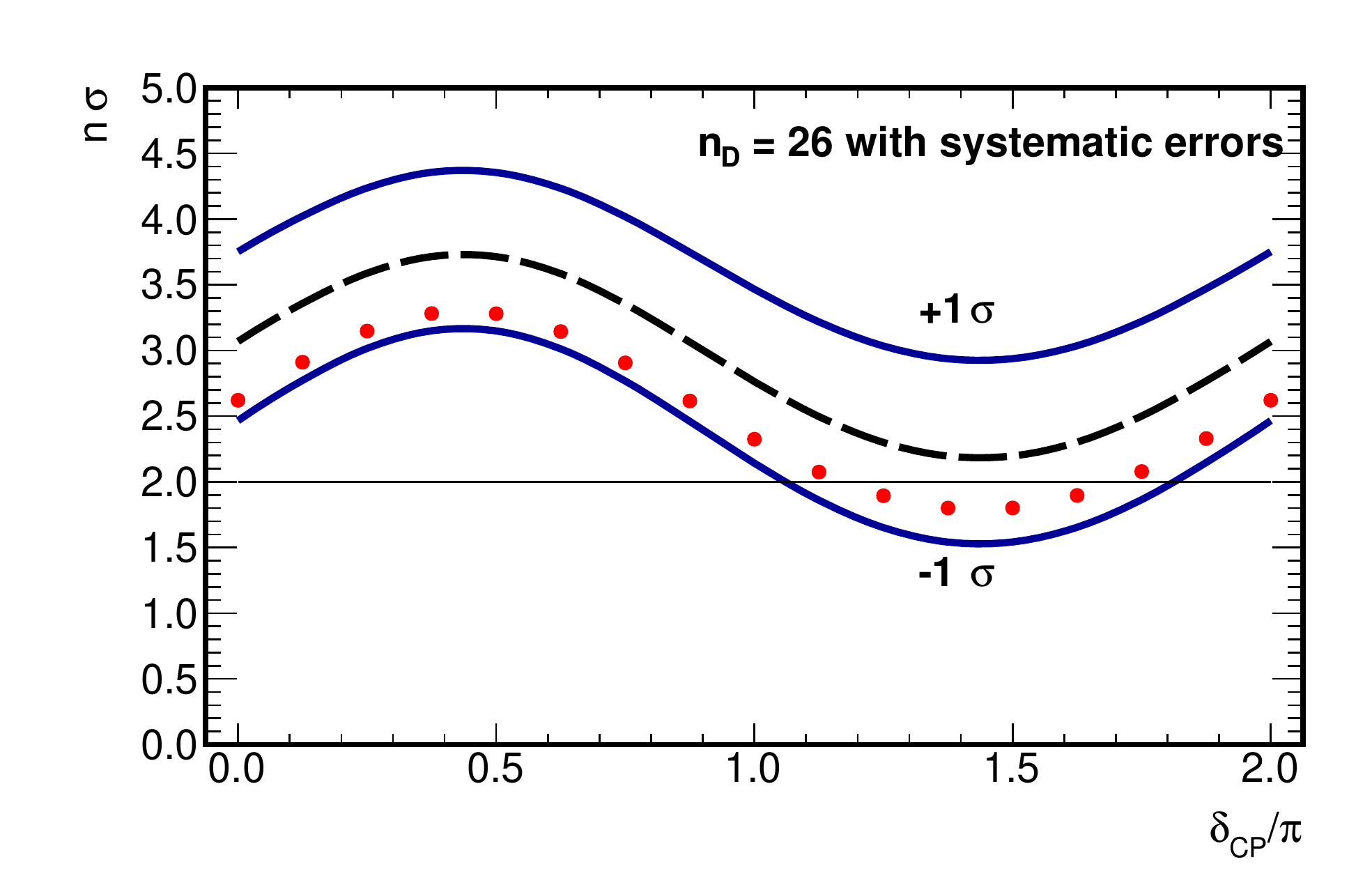}%
\caption{\label{fig5}(color online) Same as Fig.~\ref{fig4} with a total exposure of 
$8.22\times 10^{20}$ p.o.t. in NOvA, assuming the same 2015 efficiency of the signal and the same 2015 level of background rejection. 
The top picture shows the number of standard deviations for the possible observation of 18, 24 or 33 events.
Uncertainties on $\theta_{23}$, $\theta_{13}$ were included by convolution (approach A) without any systematic error.
In the bottom picture the systematic errors (approach B) of signal
and background were added to the single case $n_D=26$,
while uncertainties on $\theta_{23}$, $\theta_{13}$ were still included by convolution (approach A).
The band delimited by the two continuous lines and centered on the dashed line corresponds to the fluctuation of the significance.
When all the sources of errors are treated by convolution with the original density probability (approach A for both 
$\theta_{23}$, $\theta_{13}$ uncertainties and systematic errors) 
the significance level is shown by the dotted line. 
}
\end{figure}
In the top picture of Fig.~\ref{fig5} the significance of IH against NH hypothesis is reported (A being used for the
treatment of $\theta_{23}$, $\theta_{13}$ uncertainties). 
The 3 $\sigma$ level is  reached in the $0.2\,\pi < \delta_{CP}< 0.7\,\pi$ interval for $8\times 3=24$ events.
If 26 events are observed the IH hypothesis is rejected by more than 95\% C.L. in the full range of $\delta_{CP}$ (dashed line
in the bottom picture of Fig.~\ref{fig5}). 
Note that the 5 $\sigma$ level could also be attained, at least in a limited region of $\delta_{CP}$, when 33 events are observed
(still in the 2015 NOvA conditions). However, if NH were true, the probability to observe 33 events would be very low 
(about one per mill for the best fit values of table 1 in~\cite{lisi2016}), so indicating a tension with the 3-neutrino oscillation framework.

One should mention the decrease in significance by adding the $\theta_{23}$, $\theta_{13}$ uncertainties.
Actually, in the partial Bayesian approach where a posterior is computed from the $\theta_{23}$, $\theta_{13}$ priors the effect of their uncertainties is rather small. For the 2015 NOvA analysis on $2.74\times 10^{20}$ p.o.t.
it goes from an almost null decrease in significance
to a decrease of 0.03 $\sigma$  for observations of 6 and 11 events, respectively. The loss reaches 0.1 - 0.2 $\sigma$ when the exposure is increased by three times, i.e. $8.22\times 10^{20}$ p.o.t. still analyzed as in 2015.

\begin{figure}[htbp]
\includegraphics[width=7cm]{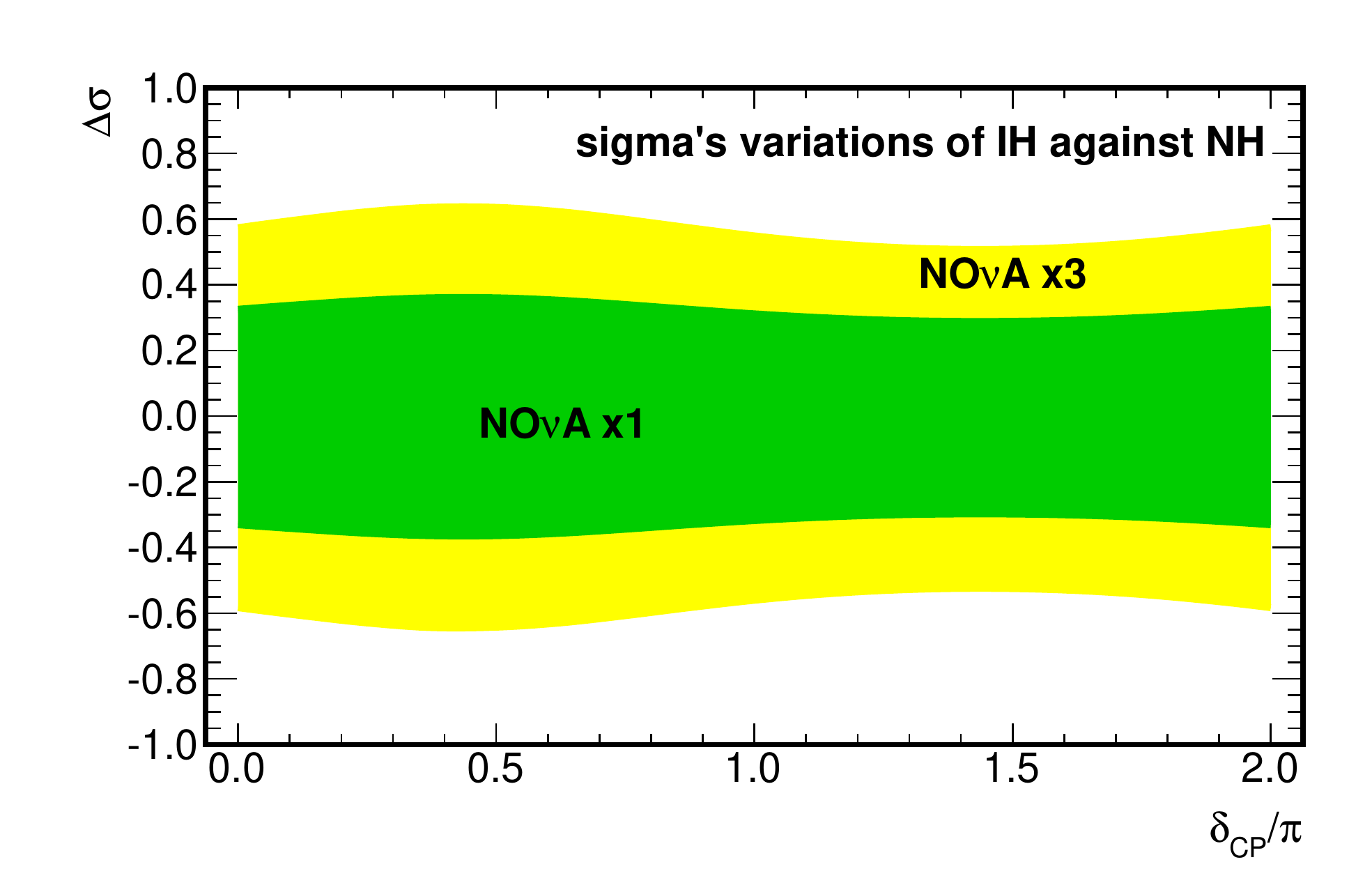}
\caption{\label{fig6}(color online) The absolute variation of the significance ($n\, \sigma$) due to the uncertainties on 
$\theta_{23}$, $\theta_{13}$, taken at 1~$\sigma$ level.
The positive correlation in the combinations of $\theta_{23}$, $\theta_{13}$ uncertainties has been considered. 
The analysis is performed for an exposure of
$2.74\times 10^{20}$ p.o.t. ($8.22\times 10^{20}$ p.o.t.) and for 8 (24) observed events, within 2015 NOvA analysis.
The result corresponds to the internal (external) band, as function of $\delta_{CP}$.
It is worthwhile to note that the loss/gain in significance is almost a factor 2 when the exposure
is three times more than in 2015, for an equivalent number of collected events and with the same kind of analysis.}
\end{figure}

The effect of adding the systematic errors with the approach B is shown in 
the bottom picture of Fig.~\ref{fig5} for the $n_D=26$ case. Including 11\%
for background expectation and 17.6\% for the signal expectation (as evaluated for the NH case and the 2015 primary selection by 
NOvA~\cite{nova-nue}) the variation of the significance 
is $\pm 0.5\, \sigma$. Instead, a loss of 0.3 -- 0.4 $\sigma$ is  obtained when all the errors are treated as nuisances  (approach A),
as reported in the same picture (dotted line).
 
\begin{figure}[htbp]
\includegraphics[width=8cm,height=6cm]{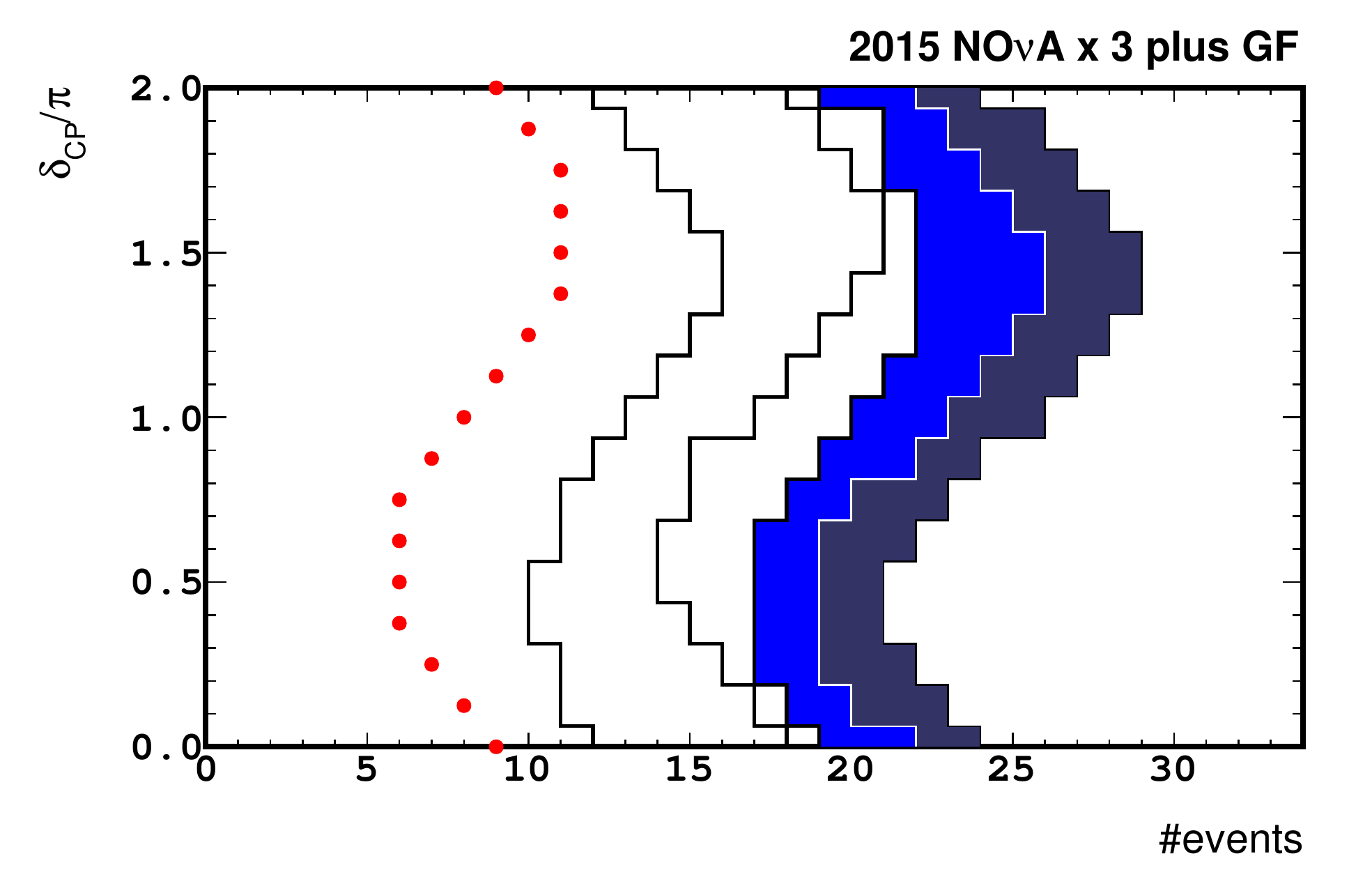}
\caption{\label{fig7}(color online) The minimum (IH exclusion, on the right) and the maximum (NH exclusion, on the left) number of events to be observed 
by NOvA for an exposure of $8.22\times 10^{20}$ p.o.t. analyzed as in 2015, are reported as function
of $\delta_{CP}$
to exclude one of the two mass ordering at 95\% C.L.. The central line inside the double-band on the right delimits
the exclusion region for IH, while 
the dotted line on the left delimits the exclusion region for the NH hypothesis.
The double-band corresponds to add $\pm 1\,\sigma$ systematic errors with approach B on the expected
signal and background (shown only for the IH exclusion case). 
The two central thick lines correspond to the median of the $f_{\rm IH}$ (left) and $f_{\rm NH}$ (right) probability densities, obtained by
the convolution of the Poisson event distribution and the Gaussian distributions for $\theta_{23}$, $\theta_{13}$ uncertainties.
}
\end{figure}

A full frequentist approach, that is B,  has also been  considered for $\theta_{23}$, $\theta_{13}$.
In this case the uncertainties on $\theta_{23}$, $\theta_{13}$ correspond to bandwidths around the median significances.
If the positive correlation of $\theta_{23}$, $\theta_{13}$ uncertainties  
is chosen at 1~$\sigma$ level, the corresponding absolute variation of the  significance is shown in Fig.~\ref{fig6}. 
An almost symmetric reduction/increase 
in the significance is observed: about 0.3 (0.6)  $\sigma$ when $q_{\rm IH}$ is computed for $n_D=8\, (24)$ 
 for a 2015 NOvA exposure of $2.74\, (8.22)\times 10^{20}$ p.o.t..

Finally, the minimum number of events that NOvA  should observe to exclude IH at 95\% C.L. is computed,
for a total exposure of $8.22\times 10^{20}$ p.o.t. analyzed as in 2015.
This is reported in Fig.~\ref{fig7}, together with the maximum number of events to exclude NH (dotted curve on the left), respectively.
For illustrative purposes the effect due to 1 $\sigma$ systematic errors (approach B) is depicted for the IH exclusion region. The median
curves of the $f_{\rm MO}$ probability densities are also drawn. If $\ge 29$ events should be observed, the new method would reject
IH at 95\% C.L., including $+1\, \sigma$ of systematic error, in the full range of $\delta_{CP}$.

\section{Discussion}\label{sec:disc}
Following the initial observation by NOvA in 2015 that mildly favours
NH and considering, for example, a three-time increase in exposure,
the new method 
based on the estimator $q_{\rm MO}$ would be able
to disfavor IH by up to 3-4 $\sigma$ depending on the $\delta_{CP}$ value. If the 2015 NOvA result,  i.e.  8-11 observed events,
should be confirmed using $8.22\times 10^{20}$ p.o.t. and about 30 events be found with the unchanged 2015 analysis,
IH could be rejected in the 3 $\nu$ framework.
The effect of the systematic errors would lower the significance
by about 0.5 $\sigma$, still sufficient to reach a firm conclusion. 
If about one third of events (i.e. about 10) would 
be observed, the NH hypothesis could be disproved at 95\% C.L.  for $\delta_{CP}>\pi$. 
If instead about 16 events would be collected
no conclusion would be possible on IH and NH over the full range of $\delta_{CP}$. 
Note that 16 events correspond to the expected averaged median
of the $f_{\rm MO}$ distributions, either for IH or NH. 
Note also that the systematic
errors reduce the gap between IH and NH expectations, pointing to the necessity of lowering them as the exposure increases.

It is relevant to outline that with the method here introduced and the treatments of the uncertainties on
$\theta_{23}$, $\theta_{13}$ and the systematic errors, a robust result can be achieved in the full range of $\delta_{CP}$
only if a moderate fluctuation, i.e. statistically acceptable,  occurs.
This conclusion sounds strange but it is consistent with the performed analysis. The repetition of the experiment (equivalent
to collecting several samples of exposure data set) will
not automatically overtake the previous result. Instead, a positive outcome can be reached when a favorable fluctuation is found.
That is detailed in the appendix in a quantitative way.

Moreover, even though statistical fluctuations are present and actually used in the analysis, 
once a result on MO is obtained (within the defined C.L., which corresponds to the correct 
coverage by construction) then the next experiment cannot reach the opposite conclusion, 
as long as both experiments handled their analyses properly.
Further, note that there is no assurance to gain more information by the second experiment, for example whether less fluctuations occur. 
This is an intrinsic property of the statistical behaviour of 
the physical process and the used estimators.

\begin{figure}[h]
\includegraphics[width=8cm]{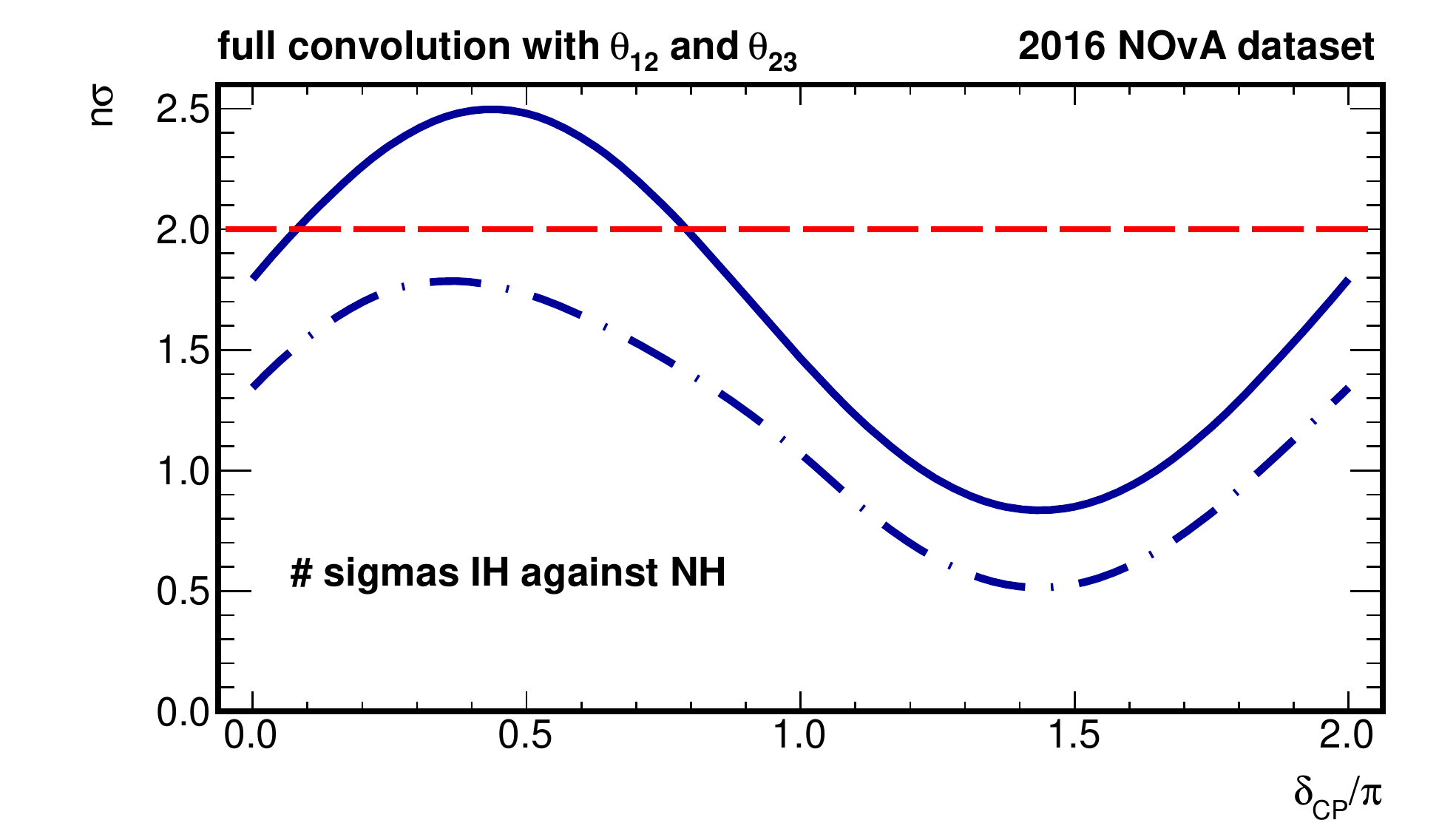}
\caption{\label{fig8}(color online) The exclusion of the inverted hierarchy as obtained by the new statistical method
applied to the recent release by NOvA in 2016~\cite{nova-prel}. Uncertainties on $\theta_{23}$, $\theta_{13}$ were included by 
convolution (approach A).
For comparison the dot-dashed line corresponds to
the $\Delta\chi^2$ result. The orizonthal dashed line indicates the 95\% C.L..
IH is rejected at 95\% C.L. in the $\delta_{CP}$ interval [0.10$\pi$, 0.77$\pi$]. 
}
\end{figure}

It is worth to look at the just released preliminary new results by the NOvA collaboration~\cite{nova-prel}.
In its update NOvA analyzed $6.05\times 10^{20}$ p.o.t., a factor 2.2 increase of the 2015 exposure.
For the \numunue appearance channel 33 events were found, including background. However, the background level was  enhanced
(a factor 4.5) against an increase in a factor 2.5 for the signal efficiency. 
By scaling these number to the 2015 analysis and exposure
the 33 events in 2016 corresponds to about 6 events in 2015. That is around the median expectation
without an even moderate fluctuation.
Anyhow, applying our new method, the increase in exposure from 2015 to 2016 allows us to obtain a first important result:
the inverted hierarchy can be excluded 
at 95\% C.L. in the $\delta_{CP}$ interval [0.10$\,\pi$, 0.77$\,\pi$] (Fig.~\ref{fig8}). We outline that the latter result is achieved
including the current $\theta_{23}$, $\theta_{13}$ uncertainties, and not fitting to their best values.

Comparison of the results for the $q_{\rm MO}$ estimator applied to the 2015 and 2016 NOvA analyses suggests
the need to carefully evaluate the contributions of signal and background to the final sample. For studies on MO  some
figures-of-merit may be more valuable than others, e.g. those used for the parameter oscillation analyses. In particular the purity
level may be more relevant than the efficiency on the signal. Moreover, a partition of the data samples may be envisaged.
Without entering in too much technical discussion the issue on blind analyses has to be considered too.

To complete the discussion, it is worthwhile to note that the foreseen NOvA run with anti--neutrinos will certainly contribute
to disentangle IH and NH, as well as adding information from the T2K experiment~\cite{t2k-nue}. 
Besides, the JUNO~\cite{juno}  measurement of MO in vacuum becomes very relevant since it will not depend
on $\delta_{CP}$.
The possible atmospheric measurements as foreseen by PINGU~\cite{pingu}  and ARCA/ORCA~\cite{km3} would contribute as well.
We plan to extend our new method here described to all these frameworks.
However, it should be clearly stated that if in the next future NOvA makes observations in line with its 2015 analysis
then the inverted hierarchy will be rejected at 95\% C.L. in the full range of $\delta_{CP}$
using the analysis reported in this paper.
Although no technical conclusion on the normal hierarchy could be possible, the logical conclusion would still be drawn since the two
hypotheses are opposite in the three-neutrino oscillation scenario.
\appendix*
\section{}

The appendix describes some characteristics of the new test statistic comparing them to the 
$\Delta\chi^2_{min}$ method. The framework of the 2015 NOvA analysis has been considered.
For simplicity only the statistical fluctuations are taken into account, neglecting
the uncertainties of the oscillation parameters $\theta_{23}$, $\theta_{13}$ and the systematic errors
of the measurements for the expected signal and background number of events.

\subsection{The standard $\chi^2$ method}

We define as $n_{\rm NH}$ ($n_{\rm IH}$) the number of predicted events in the NH (IH) hypotheses 
for the \numunue appearance at NOvA, for some hypothetical running conditions and a specified value of $\delta_{CP}$. 
Defining the variable $d = n_{\rm NH}-n_{\rm IH}$ the $\chi^2$ is
computed as  $\chi^2 = d^2/(n_{\rm NH}+n_{\rm IH})$. Its probability $P(\chi^2,1)$ for 1 d.o.f. is subsequently evaluated. 
The probability $P$ can be associated to the equivalent number $Z$ of standard deviations. Choosing the one--sided option 
$Z$ is computed as $Z=\Phi^{-1}(1-P)$, where $\Phi^{-1}$ is the quantile of the standard Gaussian distribution.

\begin{figure}[htbp]
\includegraphics[width=9cm]{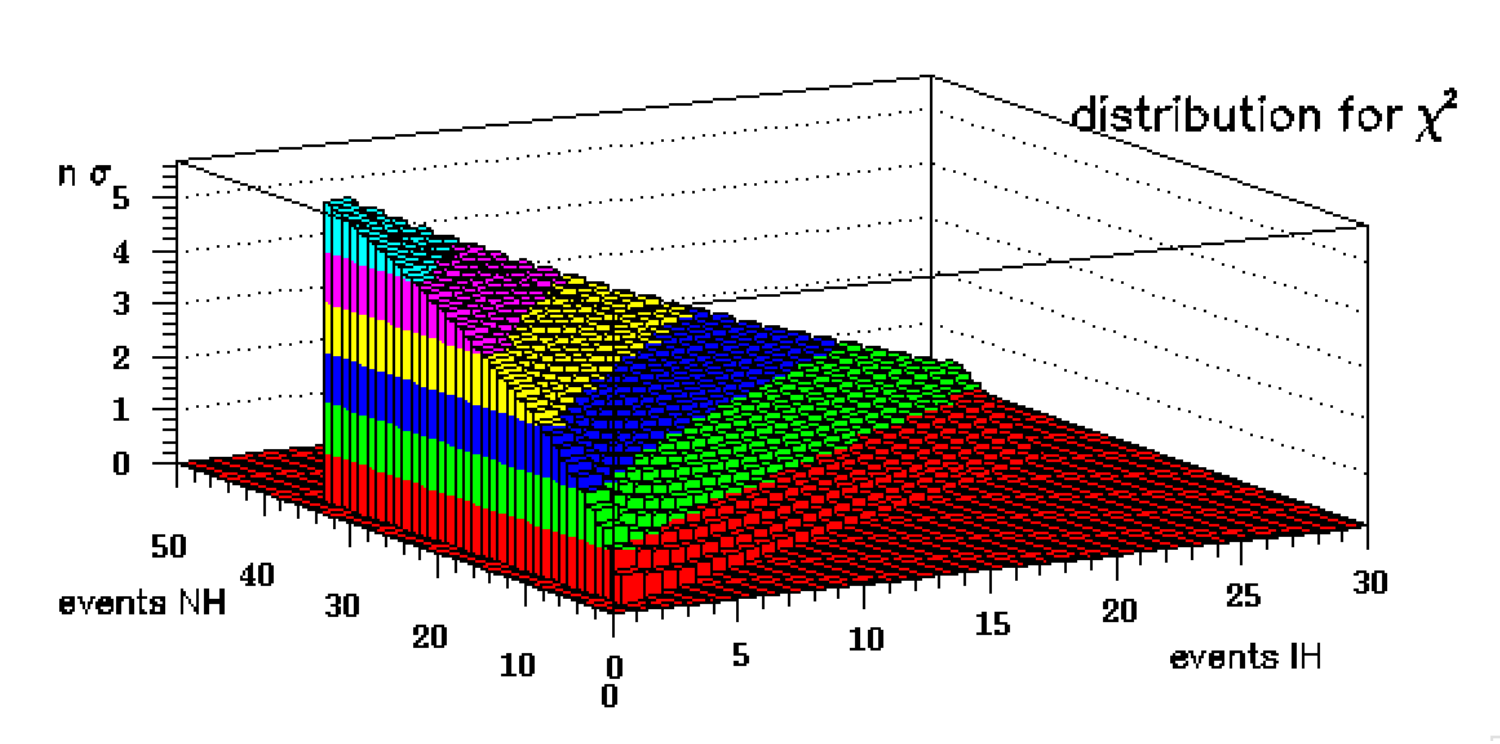}
\caption{\label{fig_a1}(color online) The number $Z$ of sigmas (as defined in the text) is drawn as function of $n_{\rm NH}$ and $n_{\rm IH}$,
the predicted numbers of events in the NH and IH hypotheses, respectively.}
\end{figure}

The number $Z$ of sigmas  is plotted in Fig.~\ref{fig_a1} as function of variables $n_{\rm NH}$ and $n_{\rm IH}$. From the plot
one estimates that e.g. when $n_{\rm IH} = 10$ are predicted in the IH hypothesis then $n_{\rm NH}$ in the NH hypothesis
should be larger than 20  events  to get a significance of 3 $\sigma$.

\begin{figure}[htbp]
\includegraphics[width=8cm]{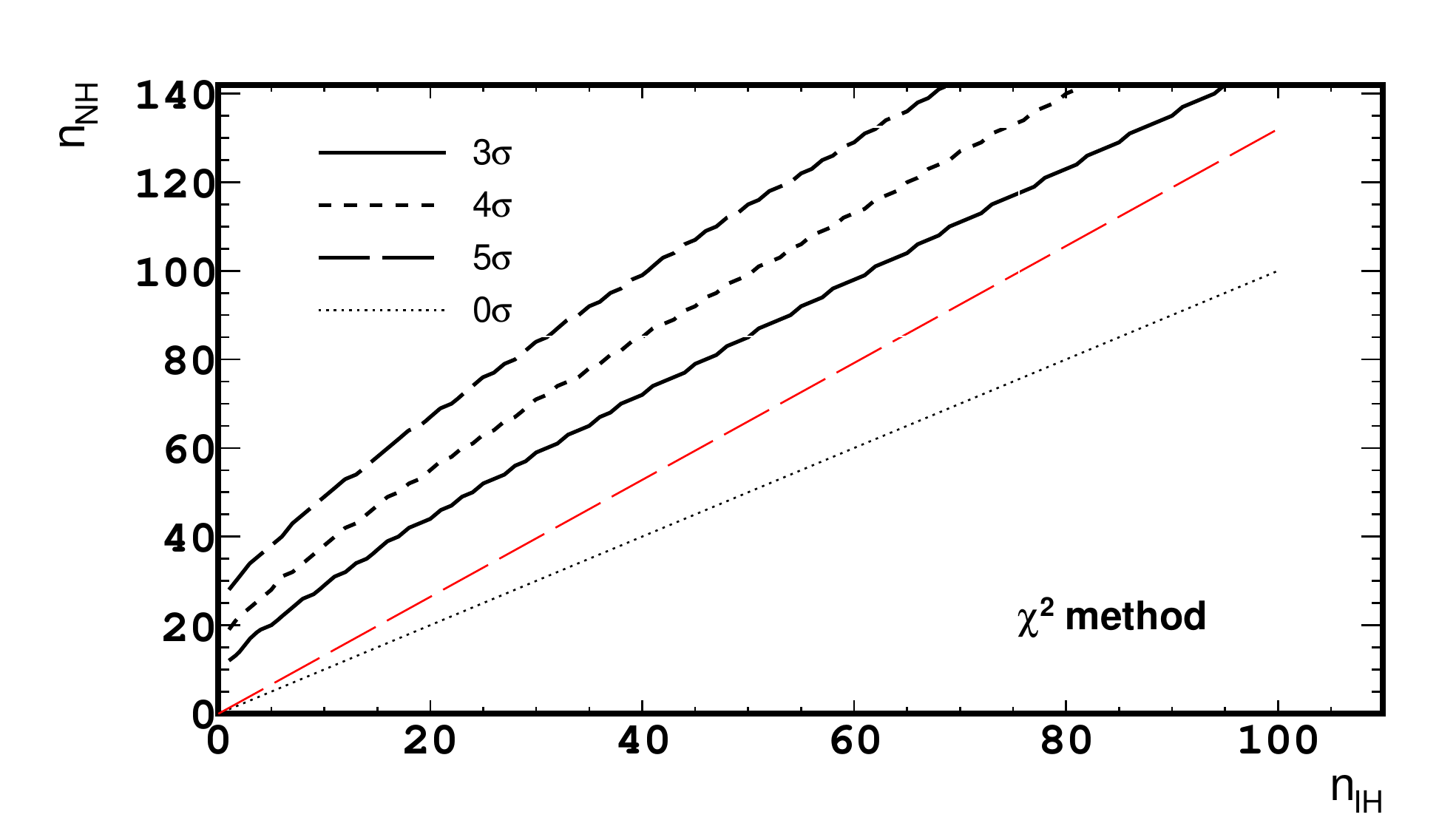}
\includegraphics[width=8cm]{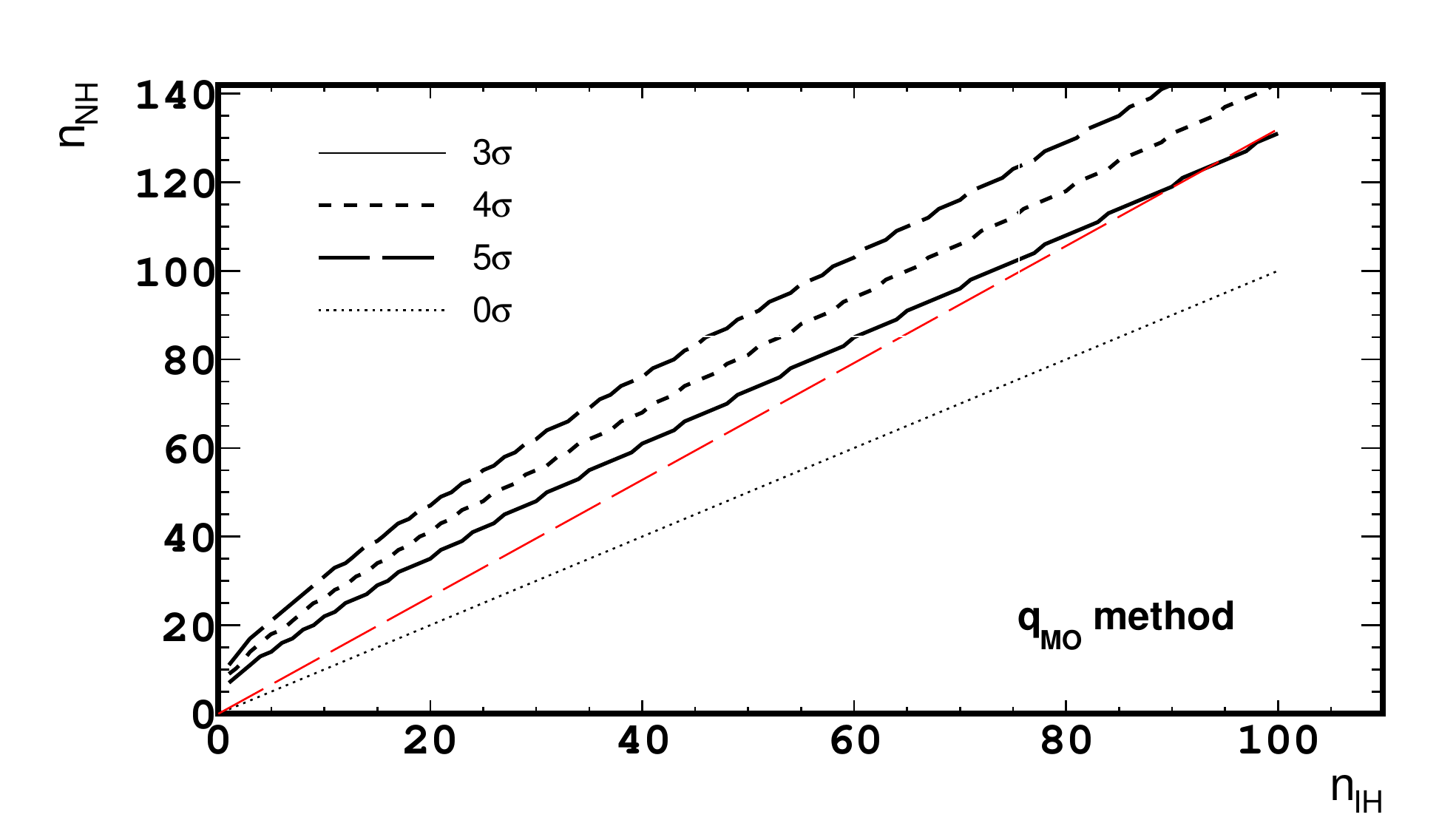}
\caption{\label{fig_a2}(color online) Using the same variables of the previous figure ($n_{\rm IH}$ on the horizontal axis
and $n_{\rm NH}$ on the vertical one) the isolines for $n\,\sigma$ significances are drawn. For example the region above
the line of $\sigma = 3$ corresponds to the combinations of ($n_{\rm IH}$, $n_{\rm NH}$) that would give more than 3 $\sigma$ significance.
The top picture shows isolines for the $\chi^2$ method, while in the bottom one isolines from the new method are drawn.
The dotted lines on the bisector correspond to zero significance being $n_{\rm NH} = n_{\rm IH}$. The dashed-red lines
immediately above 0 $\sigma$
indicate the median of the predicted number of NH events, with the 2015
NOvA conditions, $\delta_{CP}=1.5\,\pi$ and the other oscillation parameters given by the best fit of GF.
}
\end{figure}

The situation is better illustrated if isolines for given significances are computed. This is shown in Fig.~\ref{fig_a2} 
for the $\chi^2$ (top) and for the new method based on the $q_{\rm MO}$ estimator (bottom).
For example from the top picture, if 10 events are expected for IH (horizontal axis), 28 should be observed to reject IH 
(vertical axis) at 3 $\sigma$ level.
The dotted lines for 0 sigma correspond to the same number of events expected for IH and observed for NH. 
In such a case of course there is no sensitivity to distinguish IH/NH. 

In the same plots the dashed-red lines immediately above the 0 $\sigma$ isolines show the actual median expectation 
of NH in the 2015 NOvA analysis, with $\delta_{CP}=1.5\,\pi$ and the Global Fit (GF) best fit values for the other oscillation parameters.
The exposure corresponds to the number of IH events, $n_{\rm IH}$. 
Note that the background contribution has been included. Therefore, a normalization point is given by the predicted
4.28 events for IH, 5.95 events for NH, at $\delta_{CP}=1.5\,\pi$, and 0.99 events of background.
The preliminary 2016 NOvA analysis does not significantly change  the relation between IH and NH, i.e. the slope of the 
dashed-red line,
which thus depends only on $\delta_{CP}$ in this framework.

\subsection{The new estimator $q_{\rm MO}$}

In this simplified case the new test statistic $q$ is defined, for each generic $n$, as

\begin{align}
q(n) & =\frac{\sum_{n_i\ge n} Poi_{\rm IH}(n_i;n_{\rm IH})}{\sum_{n_j\ge n}
Poi_{\rm NH}(n_j;n_{\rm NH})},
\end{align}

\noindent where $Poi$ indicates the Poisson distributions with means $\mu_{\rm IH}=n_{\rm IH}$ and $\mu_{\rm NH}=n_{\rm NH}$.
Computing $q(n)$ for any $n$ and weighting them with the distribution one wants to test, e.g. $Poi({\rm IH})$,
the probability mass function $P_{\rm IH}(q(n);n_{\rm IH})$  is obtained. That is the probability distribution of $q$ under the
hypothesis that IH is the truth:
 
\begin{align}
P_{\rm IH}(q(n);n_{\rm IH}) & =Poi_{\rm IH}(n;n_{\rm IH})
\end{align}

Finally, to extract a significance for a given $n_{\rm NH}$ a $p$-value is computed :

\begin{align}
p(n_{\rm NH};n_{\rm IH}) & = \sum_{n\ge n_{\rm NH}} P_{\rm IH}(q(n)).
\end{align}

The $p$-value thus obtained, as function of $n_{\rm IH}$ and $n_{\rm NH}$, is then transformed into a significance 
by evaluating the number of standard deviations in the same way done for the $\chi^2$ probability.

The $q$ test statistic is optimal~\cite{cls} in the sense that it maximizes the probability of rejecting a false 
hypothesis, at a given confidence level, and conversely minimizes the probability 
of making a false discovery, at a given discovery confidence level. 
Comparing the two plots in Fig.~\ref{fig_a2} it is evident that $q_{\rm MO}$ is more powerful than the $\chi^2$ method. 
This intuitively originates from the
fact that the $\chi^2$ makes use of only a representative point of the distribution while the $q_{\rm MO}$ makes use
of the full information of the underlined distribution.
Then for example, instead of the 28 events needed from the $\chi^2$ to get a 3 $\sigma$ significance when 10 IH events are predicted,
only 20 are required for the $q_{\rm MO}$ test.

However,  until now the probability to observe a certain number $n_{\rm NH}$ has not been considered.
For example, the probability to observe $n_{\rm NH}=20$ when 10 events are predicted for IH should 
be looked at. To take it into account one checks when the expectation line of NH (dashed-red line
immediately above the 0 $\sigma$ line) intersects the isolines.
From the bottom plot ($q_{\rm MO}$ test) of Fig.~\ref{fig_a2} the intersection with the 3 $\sigma$ isoline occurs at about 90 events.
Instead, the $\chi^2$ test does not show any intersection in the displayed range, suggesting it will occur at rather larger $n_{\rm IH}$,
i.e. at a rather large exposure of the experiment.
It can be computed that $n_{\rm IH}=90$ corresponds to an increase in a factor 17 
of the 2015 NOvA exposure and a factor 3 in exposure of the 2016 NOvA  analysis. 
Thus, in principle, the $q_{\rm MO}$ estimator will be able to distinguish IH from NH at 3 $\sigma$ level for an increase in the above factors,
at least for $\delta_{CP}=1.5\,\pi$. However, this is a simplified case since all the error sources 
are neglected. What matter here is the relative success rate of $q_{\rm MO}$ against $\chi^2$.

One notes that the isolines tend to approach the NH prediction for both methods. This is true for all the $\delta_{CP}$ values.  
The tendency is  slow for the $\chi^2$ method, whilst is more pronounced 
for the new method. 
\begin{figure}[htbp]
\includegraphics[width=8cm]{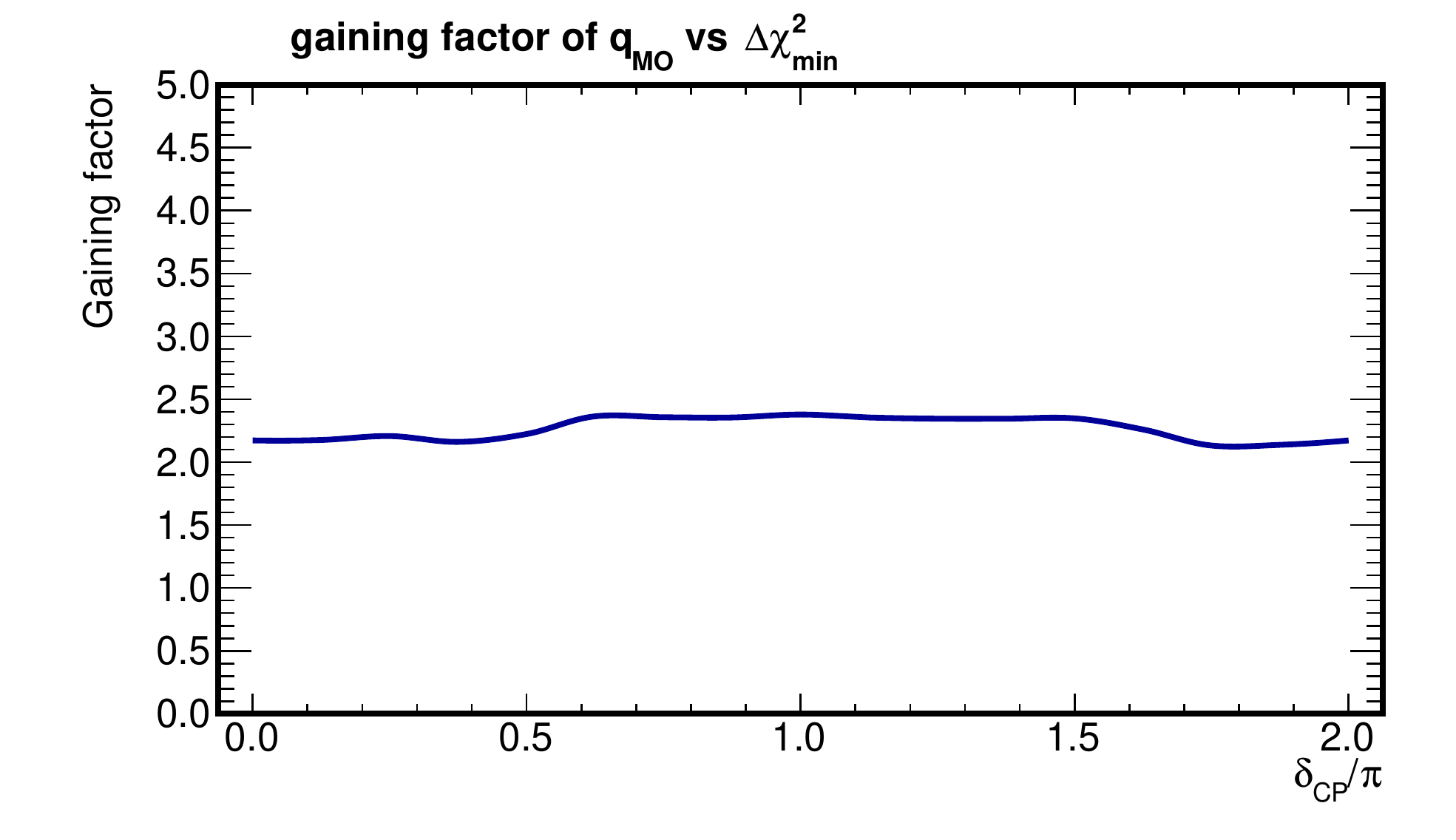}
\caption{\label{fig_a3}(color online) The improvement factor between the $\chi^2$ and the $q_{\rm MO}$ methods
to reject IH at 3 $\sigma$ level is shown as function of  $\delta_{CP}$. 
The 2015 NOvA analysis conditions have been considered for the IH/NH expectations, while the oscillation parameters
are given by best values of the global fit.
}
\end{figure}
Focussing on the  $\delta_{CP}=1.5\, \pi$ condition,
for a large number of events, i.e. for a large data sample (of the order of $2\times 10^{21}$ p.o.t.
for the 2016 NOvA analysis), there could be, in principle, the possibility to distinguish NH from IH with a significance
greater than 99\% C.L. even with the $\chi^2$ method. 
When the $q_{\rm MO}$ estimator is used about a factor two less is needed to get the 3 $\sigma$ separation. 
This corresponds to a net gain in exposure of $q_{\rm MO}$ against $\chi^2$. Such {\em gaining factor} has been quantified for each value of $\delta_{CP}$ (Fig.~\ref{fig_a3}).
The average improvement is slightly above  two. Its small increase in the $\delta_{CP}$ central region 
is due to the closer expectations 
of IH and NH (see Fig.~\ref{fig1}) where the new test statistic works even better.

This is an ideal situation that does not take into account the uncertainties on the oscillation parameters nor the systematic errors.
Even though the gain is not destroyed when errors are included in the analysis, it may take a lengthy period to collect a sufficient number of events. 
In fact, considering the uncertainties in Fig.~\ref{fig1}, the evolutions of the number of events for the two options, 
IH and NH, almost overlap.
In practice, with the current knowledge of $\theta_{23}$, $\theta_{13}$, and the current level of the systematic errors, 
there is no chance to distinguish between IH and NH in the full range of $\delta_{CP}$
neither with $\chi^2$ nor with the $q_{\rm MO}$ new method just by increasing the statistical data 
sample. 

Nevertheless, the improvement factor considerably increases when  our next
 new idea on the treatment of the data fluctuations is applied, as reported in the next section.

\subsection{Including the statistical fluctuation}

\begin{figure}[htbp]
\includegraphics[width=7cm]{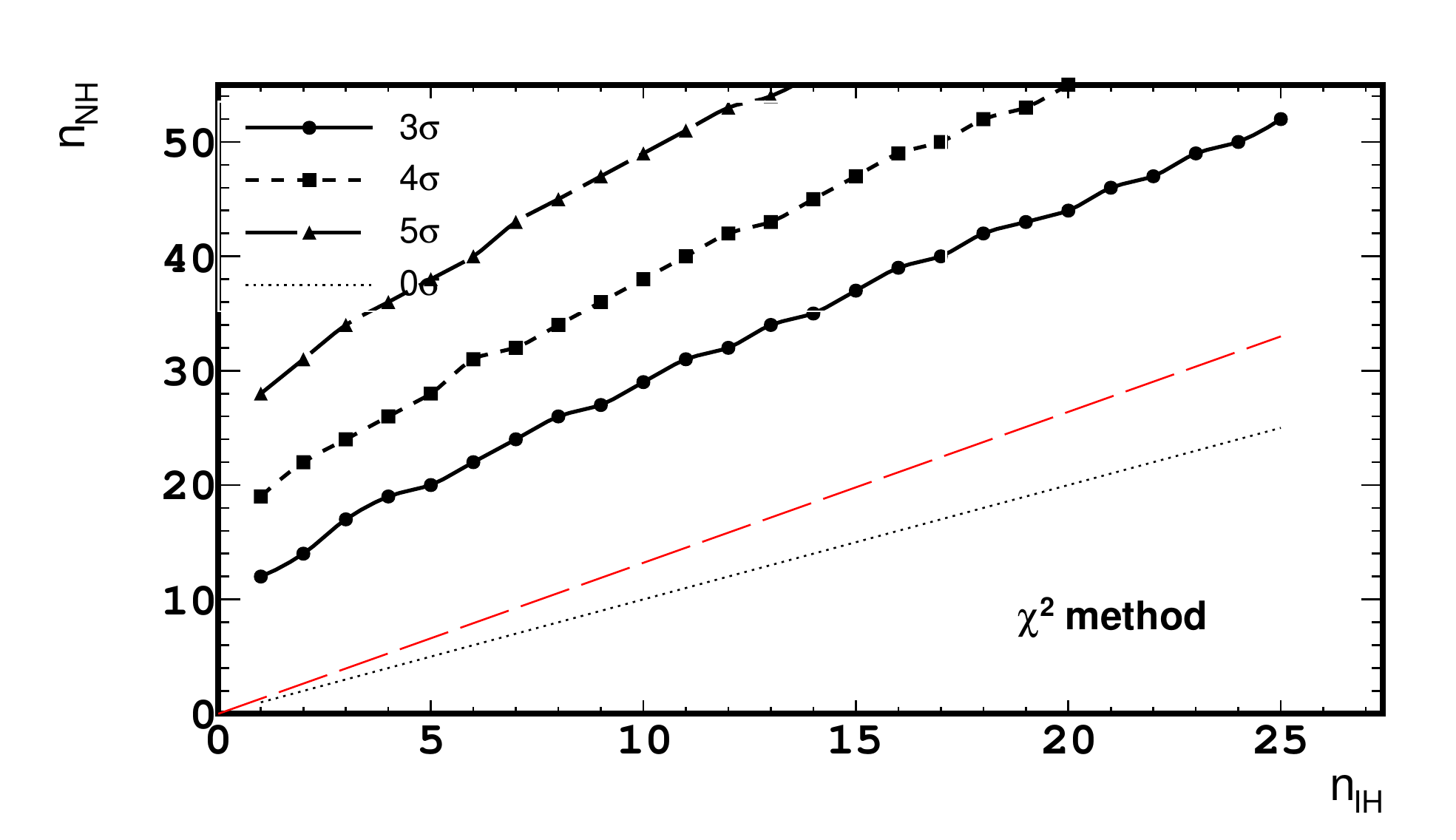}
\includegraphics[width=7cm]{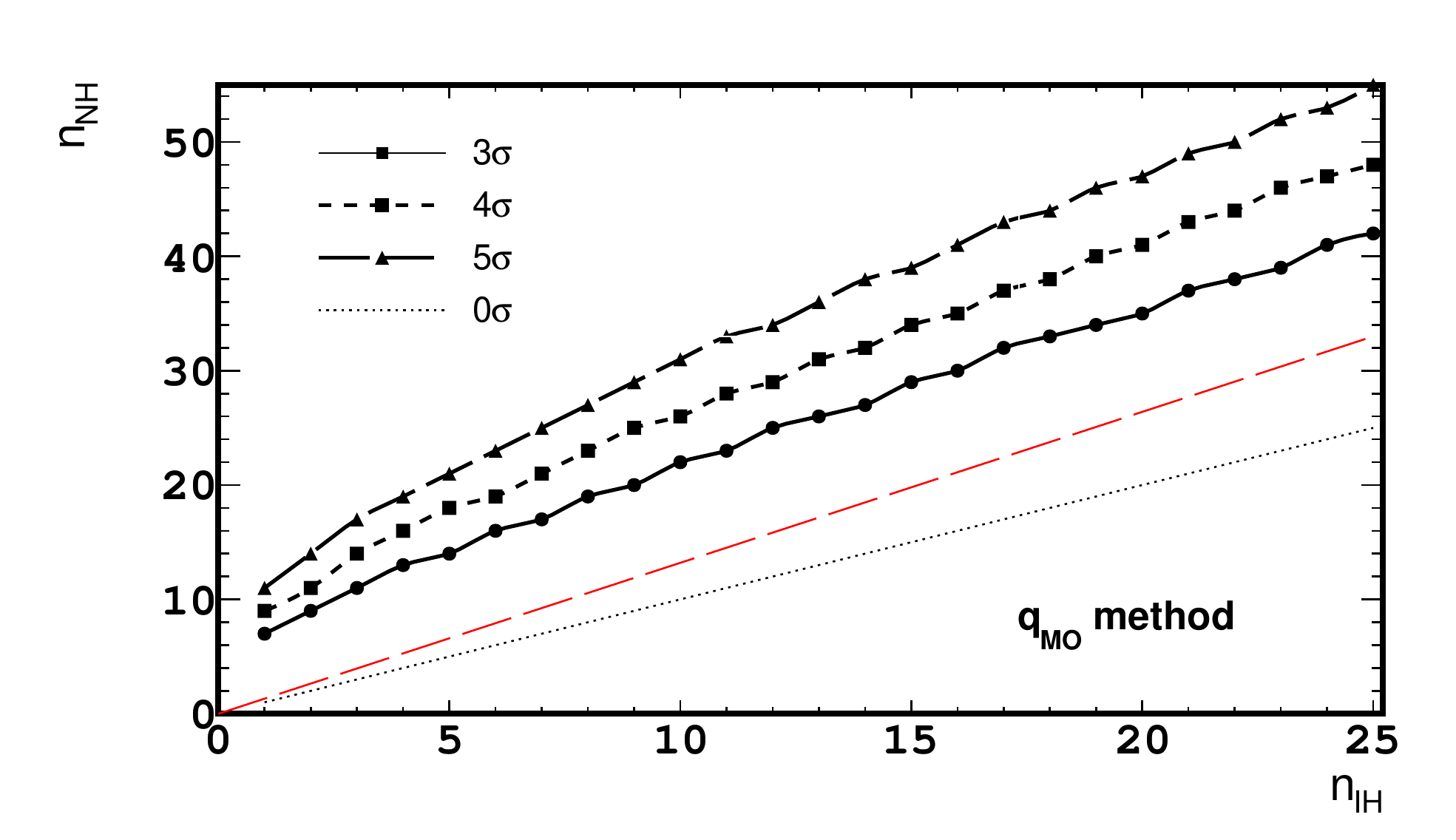}
\caption{\label{fig_a4}(color online) Same as Fig.~\ref{fig_a2}, in a restricted range of the IH number of events.
The top picture shows isolines for the $\chi^2$ method, while in the bottom one isolines for the new method are drawn.
The dotted lines on the bisector correspond to zero significance being $n_{\rm NH} = n_{\rm IH}$. The dashed-red lines
immediately above 0 $\sigma$
indicate the median of the predicted number of NH events, with the 2015
NOvA conditions, $\delta_{CP}=1.5\,\pi$ and the other oscillation parameters given by the best fit of the GF.
}
\end{figure}

Let us look at the zoomed region of $n_{\rm IH}$ in the current region of interest, $n_{\rm IH}<30$.
From Fig.~\ref{fig_a4} it is evident that even 
$q_{\rm MO}$ results
are far from the median expectation of NH in this data range. 
Thus we tried to apply the idea to allow some fluctuation of the data, mildly away from the median.
One assumes a favorable probability fluctuation around the true median (i.e. NH)
before repeating the whole computation. 

\begin{figure}[htbp]
\includegraphics[width=7cm]{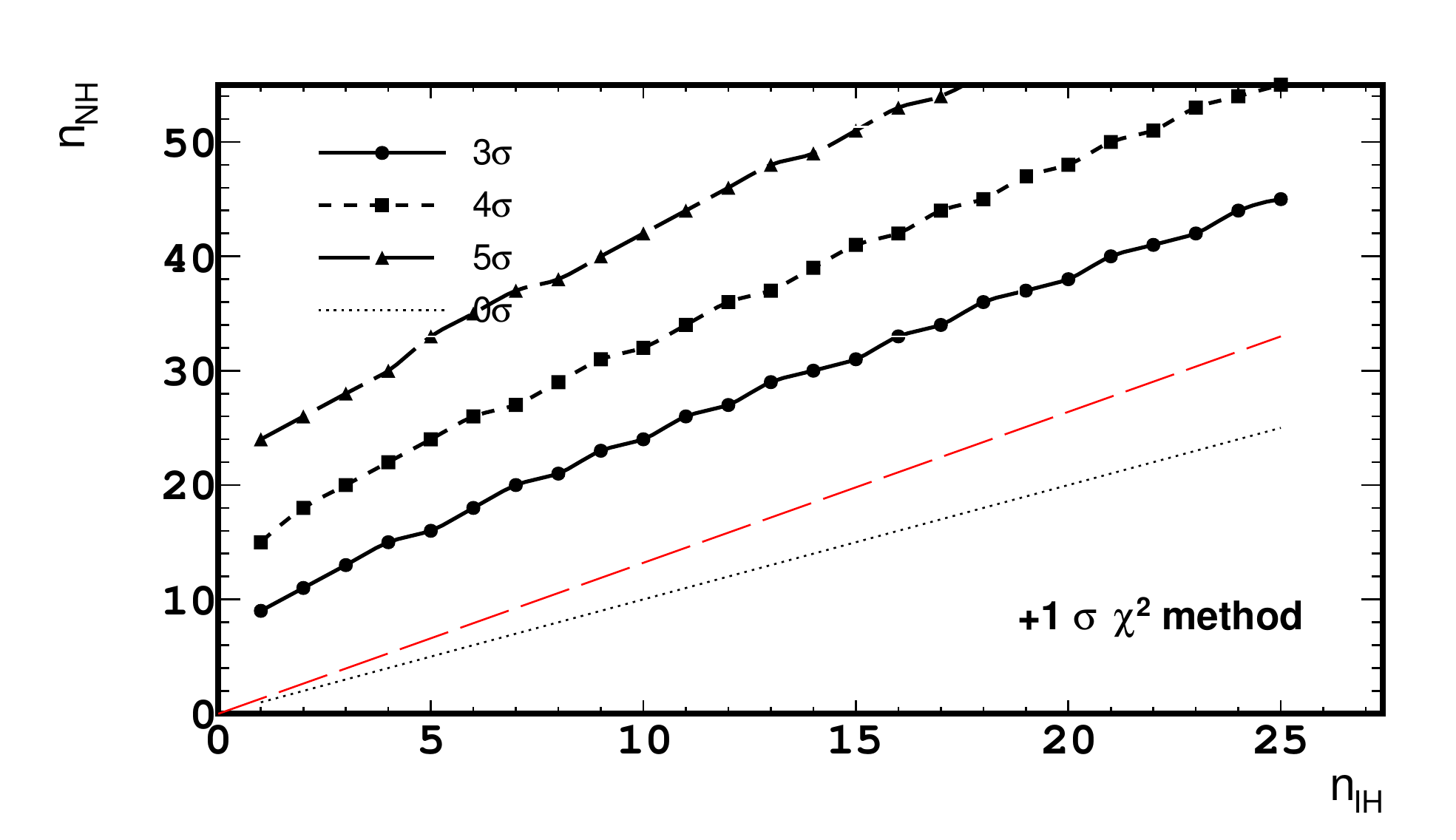}
\includegraphics[width=7cm]{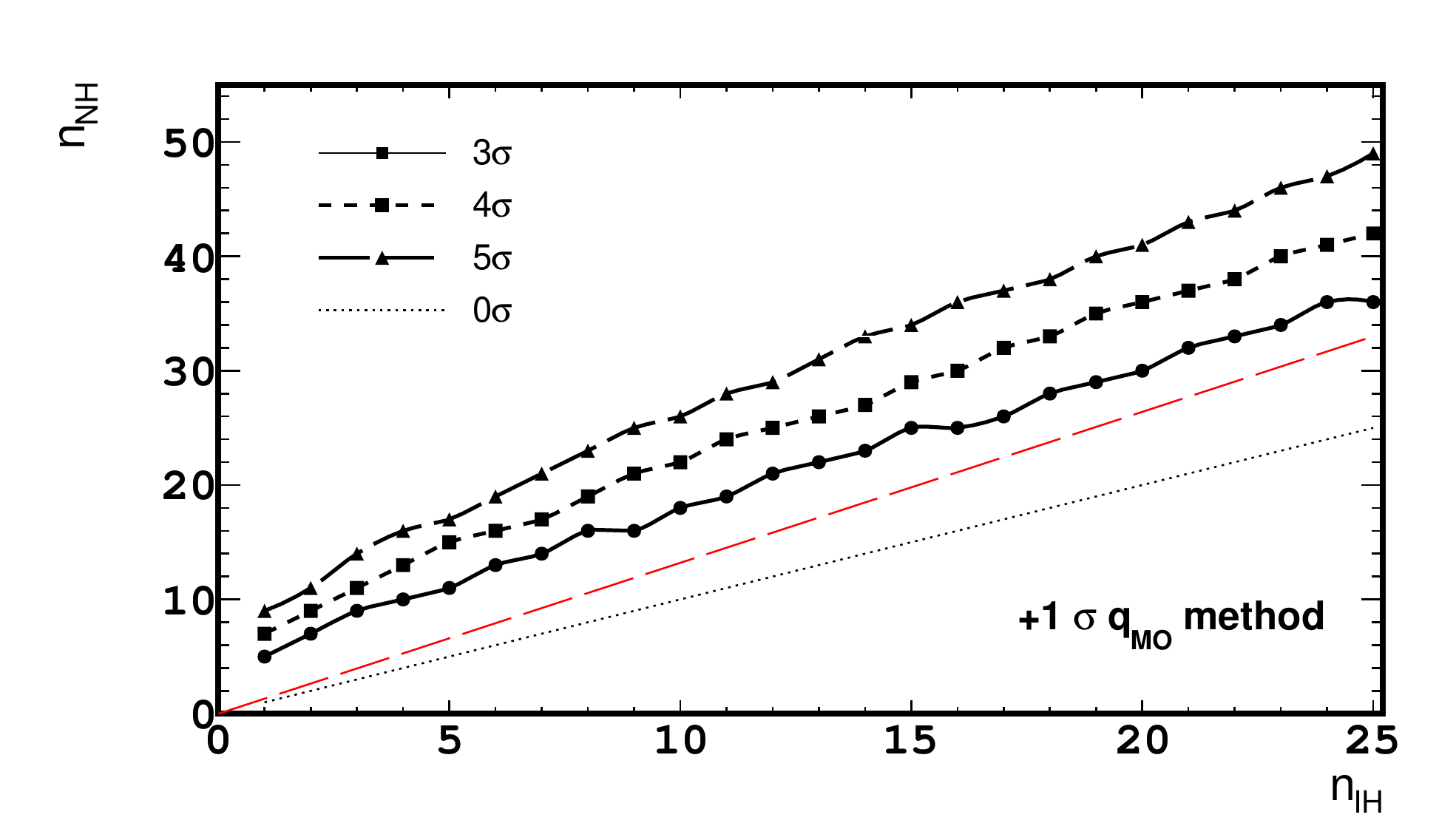}
\caption{\label{fig_a5}(color online) Isolines as in the previous figure but including a 32\% probability fluctuation on 
the number of NH events.}
\end{figure}

When  a probability fluctuation at 32\% is assumed for $n_{\rm NH}$ (approximately $n_{NH}=\mu_{NH}+\sqrt{\mu_{NH}}$),
the updated isolines are drawn in Fig.~\ref{fig_a5} for the $\chi^2$ (top) and the $q_{\rm MO}$ (bottom) methods.
Comparing plots of Fig.~\ref{fig_a4}  and Fig.~\ref{fig_a5}  some improvement is qualitatively evident  for the $\chi^2$ method, and a larger one 
 for the $q_{\rm MO}$.

\begin{figure}[htbp]
\includegraphics[width=7cm]{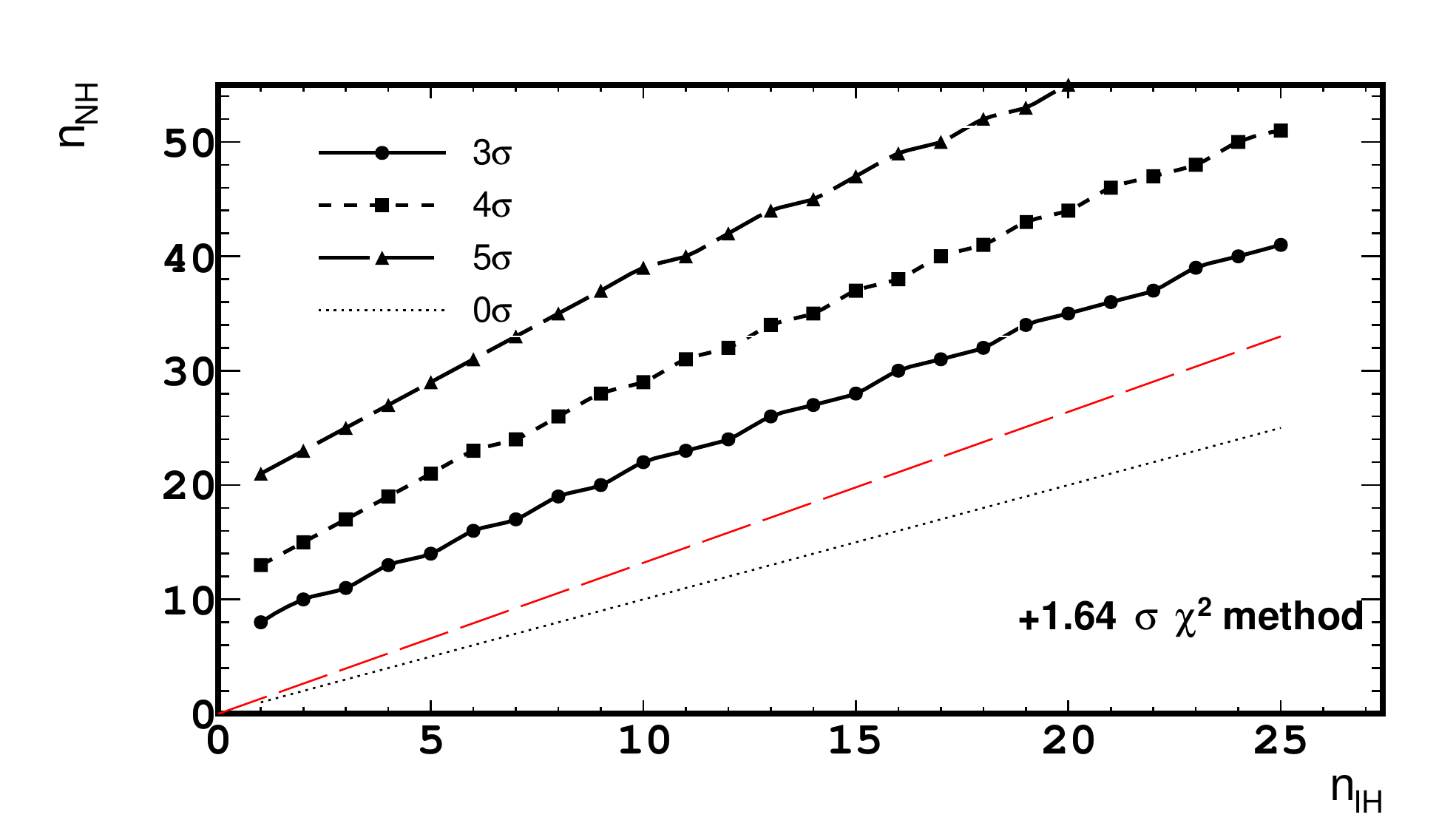}
\includegraphics[width=7cm]{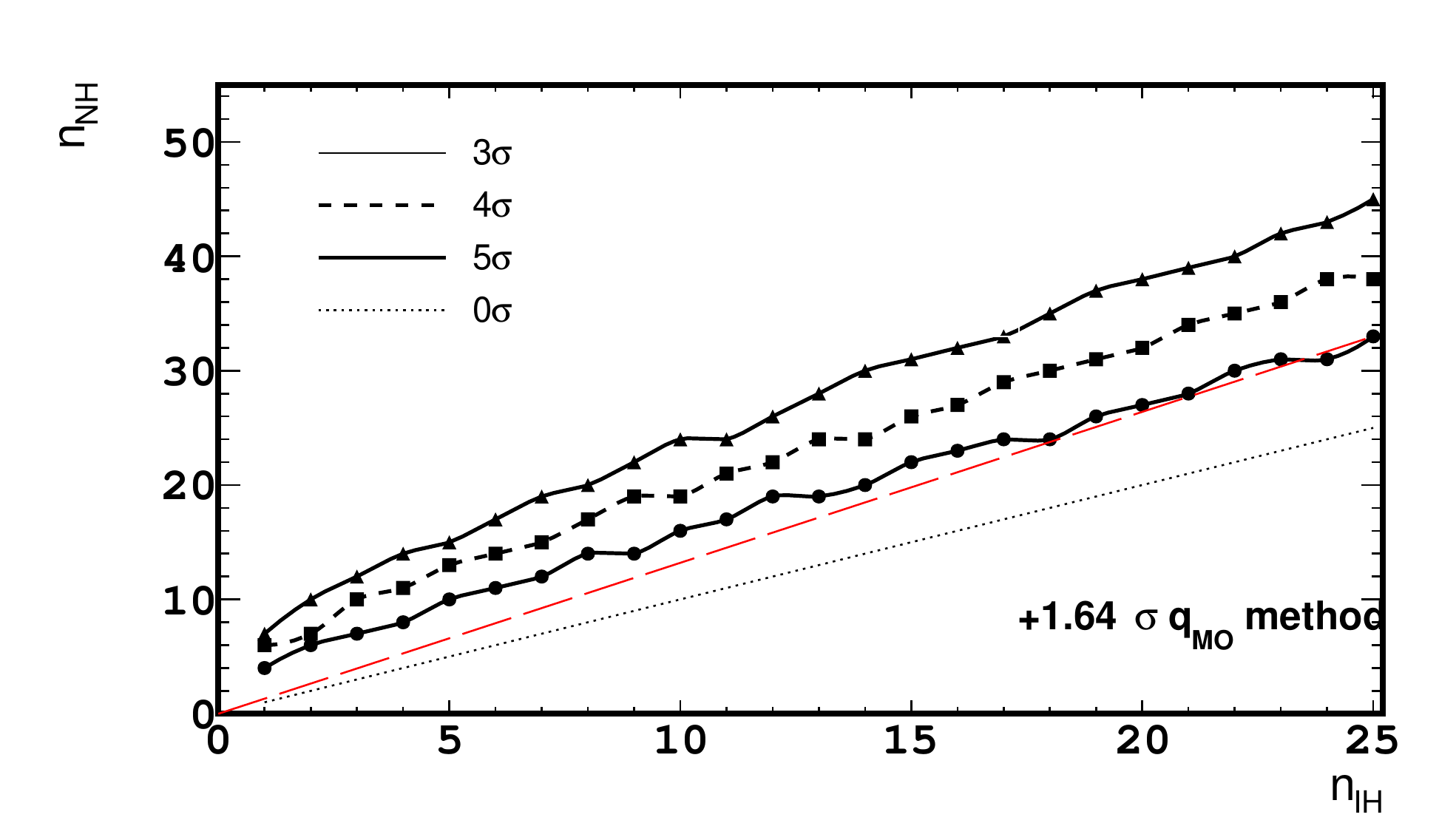}
\caption{\label{fig_a6}(color online) Isolines as in the previous figures but including a 10\% probability fluctuation on 
the number of NH events.}
\end{figure}

Results for a more pronounced  fluctuation are reported in Fig.~\ref{fig_a6}.
With a probability fluctuation at 10\% the new method allows IH to be rejected at 3 $\sigma$ level when the dataset corresponds
to about $n_{\rm}=18$. Instead, the $\chi^2$ is still far away from the possibility to put any constraint.
To be more quantitative, the gaining factors defined above have been computed for the whole $\delta_{CP}$ range. Their
averages  are reported in Tab.~\ref{tab:gaining} together with their spreads due to $\delta_{CP}$.
Note that the effect of the $\delta_{CP}$ dependence becomes more relevant when the assumed probability fluctuation increases, 
as seen from the increase in the spreads.

\begin{table}[h]
\caption{\label{tab:gaining} The gaining factors of the new $q_{\rm MO}$ method compared to the $\chi^2$ one,
for different (positive) probability fluctuations $n_{\rm NH}$. The quoted values are averaged over $\delta_{CP}$. The spreads correspond to
their maximum change in the $\delta_{CP}$  interval.}
\begin{ruledtabular}
\begin{tabular}{lcc}
fluctuation &  average & spread \\
\hline
no fluctuation &2.27 & $+ 0.18, -0.12$\\
32\% fluctuation &2.75 &$+0.51, - 0.19$ \\
10\% fluctuation & 3.78 &$+0.78, -0.39$ \\
\end{tabular}
\end{ruledtabular}
\end{table}

To conclude this is a basic demonstration that the new method works properly and is more powerful
than the standard $\chi^2$ method. More than a factor two in exposure is gained over the whole range of $\delta_{CP}$.
It becomes more powerful  (about a factor 3) when some fluctuations are observed in the data collection.
It could be the only method able to provide a significant discrimination between
IH and NH, for the current levels of uncertainties on the oscillation parameters
$\theta_{23}$, $\theta_{13}$ and  systematic errors.


\begin{thebibliography}{9}
\bibitem{ref-note} The two variables $q_{\rm MO}$ are the ratios 
of two cumulative functions, which by definition are random variables with uniform density probability in [0, 1]. Given the asymmetry 
of the Poisson distributions for IH and NH in case of the NOvA experiment, the ratios defined in equations~(\ref{eq2}) and (\ref{eq3})
stay in [0, 1] by construction.
\bibitem{check} The method does not really need the double definition of $q_{\rm IH}$ and $q_{\rm NH}$ since they are complementary. 
All the results can be obtained using only one expression and taking the $p$-value in the proper domain for either IH or NH. We defined the procedure
described in the text at ease of the reader.
\end{thebibliography}
\end{document}